\documentclass[final,5p,times,twocolumn]{elsarticle}

\usepackage{graphicx}

\usepackage{amssymb}
\usepackage{amsthm}

\usepackage{amsmath,url}

\usepackage{algorithm}
\usepackage{algorithmic}
\usepackage{pdflscape}

  \newcommand{\norm}[2][]{\ensuremath{\left\|#2\right\|_{#1}}}

  \newcommand{\abs}[2][]{\ensuremath{\left|#2\right|}}

  \newcommand{\pad}[2][]{\ensuremath{\frac{\partial #1}{\partial #2}}}

\newtheorem{definition}{Definition}
\newtheorem{problem}{Problem}

\journal{Computer Physics Communications}

\begin{document}

\begin{frontmatter}



\title{Achievable Efficiency of Numerical Methods for Simulations of Solar Surface Convection}


\author[fmv]{H.~Grimm-Strele\corref{cor1}\fnref{now}} 
\ead{hannes.grimm-strele@univie.ac.at} 
\author[fmv]{F.~Kupka} 
\author[fmv]{H.~J.~Muthsam}

\cortext[cor1]{corresponding author}
\fntext[now]{Present address: Max--Planck Institute for Astrophysics, 
             Karl--Schwarzschild--Strasse~1, D-85748~Garching, Germany}
\address[fmv]{Institute of Mathematics, University of Vienna, 
              Oskar--Morgenstern--Platz 1,
              A-1090~Vienna, Austria}

\begin{abstract}
  We investigate the achievable efficiency of both the time and the space 
  discretisation methods used in Antares for mixed parabolic--hyperbolic
  problems. We show that the fifth order variant of WENO combined with a
  second order Runge--Kutta scheme is not only more accurate than standard
  first and second order schemes, but also more efficient taking the 
  computation time into account. 
  Then, we calculate the error decay rates of WENO with several explicit 
  Runge--Kutta schemes for advective and diffusive problems with smooth and 
  non-smooth initial conditions. With this data, we estimate the computational
  costs of three-dimensional simulations of stellar surface convection and
  show that SSP\,RK(3,2) is the most efficient scheme considered in this
  comparison.
\end{abstract}


\end{frontmatter}

The simulation code Antares \citep{MuthsamKupkaLoew-Basellietal2010} was 
developed for the simulation of solar and stellar surface convection. Recently
it has also been applied to many other astrophysical
problems \citep[e.g.][]{MundprechtMuthsamKupka2013,ZaussingerSpruit2013}.

In this code, the Navier--Stokes equations (usually without magnetic field) 
and with radiative transfer (radiation hydrodynamics, RHD) are solved in the
form

\begin{subequations}
\begin{align}
  \pad[\rho]{t} + \nabla \cdot \left( \rho {\bf u} \right) & = 0, \\
  \pad[\left( \rho {\bf u} \right)]{t} 
    + \nabla \cdot \left( \rho {\bf u} \otimes {\bf u} \right) + \nabla p 
  & = \rho {\bf g} +  \nabla \cdot \tau, \\
  \pad[E]{t} + \nabla \cdot \left( {\bf u} \left( E + p \right) \right)
  & = \rho \left( {\bf g} \cdot {\bf u} \right) 
    + \nabla \cdot \left( {\bf u} \cdot \tau \right) + Q_{\rm rad}.
\end{align}\label{eq-NS}
\end{subequations}

The meaning and units of all variables is shown in Table~\ref{tab-variables}. 
An equation of state must be specified to complete this set of equations. The
viscous stress tensor $\tau = \left( \tau_{i,j} \right)_{i=1,2,3}$ is given by

\begin{equation}
  \tau_{i,j} = \eta \left( \pad[u_i]{x_j} + \pad[u_j]{x_i} 
        - \frac{2}{3} \delta_{i,j} \left( \nabla \cdot {\bf u} \right) \right)
             + \zeta \, \delta_{i,j} \left( \nabla \cdot {\bf u} \right). 
  \label{eq-tau}
\end{equation}

${\bf g}$ is the gravity vector and $Q_{\rm rad}$ is the radiative heating
rate describing the energy exchange between gas and radiation. $\delta_{i,j}$
is the Kronecker symbol. $\eta$ and $\zeta$ are the first and second 
coefficients of viscosity. 

\begin{table}[ht]
\centering
\begin{tabular}{p{0.15\columnwidth}|p{0.45\columnwidth}|p{0.2\columnwidth}}
  variable & meaning & unit (CGS) \\
\hline
  $\rho$        & gas density                        & ${\rm g}\,{\rm cm}^{-3}$                                \\
  $T$           & temperature                        & ${\rm K}$                                               \\
  $p$           & pressure                           & ${\rm dyn}\,{\rm cm}^{-2}$                              \\
  $u$           & $x$ velocity (vertical)            & ${\rm cm}\,{\rm s}^{-1}$                                \\
  $v$           & $y$ velocity (horizontal)          & ${\rm cm}\,{\rm s}^{-1}$                                \\
  $w$           & $z$ velocity (horizontal)          & ${\rm cm}\,{\rm s}^{-1}$                                \\
  $Q_{\rm rad}$ & radiative heating rate             & ${\rm erg}\,{\rm s}^{-1}\,{\rm cm}^{-3}$                \\
  $v_{\rm snd}$ & sound speed                        & ${\rm cm}\,{\rm s}^{-1}$                                \\
  $E$           & total energy                       & ${\rm erg}\,{\rm cm}^{-3}$                              \\
  $e$           & internal energy                    & ${\rm erg}\,{\rm cm}^{-3}$                              \\
  $\epsilon$    & specific internal energy           & ${\rm erg}\,{\rm g}^{-1}$                               \\
  $\eta$        & dynamic viscosity                  & ${\rm g}\,{\rm cm}^{-1}\,{\rm s}^{-1}$                  \\
  $\zeta$       & second (bulk) viscosity            & ${\rm g}\,{\rm cm}^{-1}\,{\rm s}^{-1}$                                               
\end{tabular}
\caption{Variable names, meaning and CGS units as used in this paper. Note 
         that $x$ denotes the vertical direction. Vectors are written in bold
         face. The velocity vector is ${\bf u} = \left( u, v, w \right)^T$.}
         \label{tab-variables}
\end{table}

We can rewrite equations~\eqref{eq-NS} as

\begin{subequations}
\begin{equation}
  \pad[{\bf Q}]{t} + \nabla \cdot {\bf F}_{\rm adv} 
  = \nabla \cdot {\bf F}_{\rm visc} + {\bf S}
\end{equation}

\noindent with

\begin{equation}
\begin{split}
  {\bf Q} = \left( \begin{array}{c}
                     \rho         \\
                     \rho {\bf u} \\
                     E            \\
                   \end{array} \right),\ &
  {\bf F}_{\rm adv}  = \left( \begin{array}{c}
                                \rho {\bf u}                               \\
                                \rho {\bf u} \otimes {\bf u} + p\,{\rm Id} \\
                                {\bf u} \left( E + p \right)               \\
                              \end{array} \right), \\
  {\bf F}_{\rm visc} = \left( \begin{array}{c}
                               0                  \\
                               \tau               \\
                               {\bf u} \cdot \tau \\
                              \end{array} \right),\ &
  {\bf S} = \left( \begin{array}{c}
                     0                                                       \\
                     \rho {\bf g}                                            \\
                     \rho \left( {\bf g} \cdot {\bf u} \right) + Q_{\rm rad} \\
                   \end{array} \right).
\end{split}
\end{equation} \label{eq-NScons}
\end{subequations}

${\bf Q}$ is the vector containing the conserved quantities and ${\rm Id}$ is 
the identity matrix. We call the terms collected in ${\bf F}_{\rm adv}$ the 
{\it advective} or {\it inertial} part and in ${\bf F}_{\rm visc}$ the {\it
viscous} part of the Navier--Stokes equations. All first derivatives are
contained in $\nabla \cdot {\bf F}_{\rm adv}$, all second order terms in 
$\nabla \cdot {\bf F}_{\rm visc}$. We note that $\pad[{\bf Q}]{t} + \nabla 
\cdot {\bf F}_{\rm adv} = 0$ is of hyperbolic type, whereas $\pad[{\bf Q}]{t} 
- \nabla \cdot {\bf F}_{\rm visc} = 0$ is a parabolic system.

\section{Discretisation and Numerical Methods}
\label{sec-methods}

Following the {\it method of lines} approach of discretising space and time 
separately \citep{Toro2009,LeVeque2007}, equations~\eqref{eq-NScons}
are discretised in space only and converted to

\begin{equation}
  \pad[{\bf Q}]{t} = \mathrm{L} \left( {\bf Q} \right), \label{eq-semidisc}
\end{equation}

\noindent where $\mathrm{L}$ is the operator resulting from the spatial 
discretisation of $- \nabla \cdot {\bf F}_{\rm adv} + \nabla \cdot 
{\bf F}_{\rm visc} + {\bf S}$. 
In principle, the integration of this equation can be performed with any 
numerical method for solving ordinary differential equations, in particular
Runge--Kutta methods, provided they are numerically stable, although further
properties (such as positivity, e.g., of $T$ or $E$) may be required (cf., for
instance, \citet{KupkaHappenhoferHiguerasKoch2012}).

The spatial discretisation is done separately for ${\bf F}_{\rm adv}$ and 
${\bf F}_{\rm visc}$ as defined in equations~\eqref{eq-NScons}.
In optically thin regions the radiative heating rate $Q_{\rm rad}$ is a source 
term and is calculated separately by the radiative transfer solver as 
described in~\citet{MuthsamKupkaLoew-Basellietal2010}. In optically thick
regions, the diffusion 
approximation

\begin{equation}
  Q_{\rm rad} = \nabla \cdot \left( \kappa \nabla T \right)
\end{equation}
 
\noindent is valid such that we can include $Q_{\rm rad}$ in the ${\bf F}_{\rm visc}$ term.

For ${\bf F}_{\rm adv}$, the WENO finite difference scheme is employed 
\citep{ShuOsher1988,Shu2003,Merriman2003}. The WENO scheme is a highly 
efficient shock-capturing scheme which we consider here in its fifth order
variant called WENO5. In the context of solar surface convection simulations,
its superiority in terms of accuracy compared to other high-order schemes was
shown in \citet{MuthsamLoew-BaselliOberscheideretal2007}. Its main part, the
fifth order accurate reconstruction operator, is summarised in
Algorithm~\ref{alg-weno}.

For ${\bf F}_{\rm visc}$, the fourth-order accurate scheme 
from~\citet{HappenhoferGrimm-StreleKupkaetal2013} is used. First, we outline 
the procedure for the one-dimensional diffusion equation

\begin{equation}
  \pad[\phi]{t} - D \frac{\partial^2 \phi}{\partial x^2} = 
  \pad[\phi]{t} - \pad{x}\left( D \pad[\phi]{x} \right) = 0
\end{equation}

\noindent with the constant coefficient of diffusion $D$. In one spatial 
dimension and on an equidistant Cartesian grid, the outer derivative is
approximated by

\begin{subequations}
\begin{equation}
  \pad{x}\left(\pad[\phi]{x}\right)\left(x_{i}\right)
  = \frac{\pad[\phi]{x}\left(x_{i+\frac{1}{2}}\right) 
  - \pad[\phi]{x}\left(x_{i-\frac{1}{2}}\right)}{\delta x}
\end{equation}

\noindent with constant grid spacing $\delta x$. Then, the inner derivative is 
calculated by 

\begin{equation}
  \pad[\phi]{x}\left(x_{i-\frac{1}{2}}\right) = 
  \frac{\phi_{i-2} - 15 \phi_{i-1} + 15 \phi_{i} - \phi_{i+1}}{12\,\delta x},
  \label{eq-drvb}
\end{equation} \label{eq-diffstencils}
\end{subequations}

\noindent leading to a fourth-order accurate approximation. Here, $\phi_i 
= \phi \left( x_i \right)$.

Similar procedures can be applied to any second-order term, in particular to 
${\bf F}_{\rm visc}$. Special care has to be taken for mixed derivatives. In 
the two-dimensional case and considering only the ${\bf F}_{\rm visc}$ terms, 
we arrive at 

\begin{equation}
  \pad{t} \left( \rho u \right)  = \pad{x} \left( \left( \zeta + \frac{4}{3} \eta \right) \pad[u]{x}
                                                + \left( \zeta - \frac{2}{3} \eta \right) \pad[v]{y} \right)
                                 + \pad{y} \left( \eta \left( \pad[u]{x}  + \pad[v]{y} \right)       \right)
  \label{eq-Fvis}
\end{equation}

\noindent by virtue of equations~\eqref{eq-NS} and~\eqref{eq-tau}. The outer 
derivatives are replaced by a finite difference, evaluating the inner function
at the half-integer nodes. Therefore, we need the terms inside the spatial
derivatives in~\eqref{eq-Fvis} at $(i-\frac{1}{2},j)$ and at
$(i,j-\frac{1}{2})$. $\pad[u]{x}$ at $(i-\frac{1}{2},j)$ and $\pad[v]{y}$ at
$(i,j-\frac{1}{2})$ can be calculated directly by formula~\eqref{eq-drvb}. 
Then, the coefficient functions must be interpolated to the half-integer grid.
To fourth-order accuracy,

\begin{equation}
  \eta_{i-\frac{1}{2},j} = \frac{ - \eta_{i-2,j} + 7 \eta_{i-1,j} + 7 \eta_{i,j} - \eta_{i+1,j}}{12},
  \label{eq-cbvx}
\end{equation}

\noindent assuming that the variable is given as a cell average. To calculate 
$\pad[v]{y}$ at the half integer index $\left( i-\frac{1}{2}, j \right)$, we
calculate the derivative at the cell centre by

\begin{equation}
  \pad[v]{y}|_{i,j} = \frac{v_{i,j-2} - 8 v_{i,j-1} + 8 v_{i,j+1} - v_{i,j+2}}{12\,\delta y},
  \label{eq-drvc}
\end{equation}

\noindent and then interpolate the result to $\left( i-\frac{1}{2}, j \right)$ 
according to formula~\eqref{eq-cbvx}. The computation of $\pad[u]{x}$ at 
$\left( i, j-\frac{1}{2} \right)$ is done analogously. The resulting procedure
is fourth-order accurate.

After the spatial discretisation step, the equations~\eqref{eq-NS} are 
transformed to the form~\eqref{eq-semidisc}. Since~\eqref{eq-semidisc} is an
ordinary differential equation, we can use Runge--Kutta schemes to integrate 
it.
 
We follow \citet{GottliebShuTadmor2001} in defining some basic properties of 
Runge--Kutta schemes.

\begin{definition}
Let an initial value problem of the form

\begin{equation}
  \phi'(t) = \mathrm{L} \left( \phi(t) \right),\ \phi(0) = \phi_0,
\end{equation}

\noindent be given. An {\it explicit $s$--stage Runge--Kutta} scheme is an 
integration scheme of the form

\begin{equation}
\begin{split}
  \phi^{(0)} & = \phi^n, \\
  \phi^{(i)} & = \sum_{k=0}^{i-1} \left( \alpha_{i,k}\,\phi^{(k)} + 
                   \delta t\,\beta_{i,k}\,\mathrm{L}(\phi^{(k)}) \right),\ 
                                     \alpha_{i,k} \geq 0,\ i=1,\hdots,s, \\
  \phi^{n+1} & = \phi^{(s)},
\end{split} \label{eq-erkshuosher}
\end{equation}

\noindent where $\phi^n = \phi(t_n)$ and the time step $\delta t$ is given 
by the CFL condition.
\end{definition}

\begin{definition}
Assume that $\mathrm{L}$ results from the discretisation of a spatial operator 
and let a seminorm $\norm{ \cdot }$ be given. Following 
\citet{WangSpiteri2007}, a Runge--Kutta method of the
form~\eqref{eq-erkshuosher} is called {\it strong stability preserving (SSP)} 
if for all stages $i$, $i=1, 2, \hdots s$,

\begin{equation}
  \norm{\phi^{(i)}} \leq \norm{\phi^n} \label{eq-ssp}
\end{equation}

\noindent with a CFL restriction on the time step $\delta t$.
\end{definition}

The {\it total variation diminishing (TVD)} property \citep{ShuOsher1988} is a 
special case of this definition. It results from inserting the {\it total
variation} norm of $\phi$ at time $t_n$,

\begin{equation}
  \mathrm{TV}(\phi^n) = \sum_j \abs{\phi_{j+1}^n - \phi_j^n},
\end{equation}

\noindent in~\eqref{eq-ssp}.

In this paper, we consider four explicit time integration schemes: the 
first-order Euler forward method, the second-order two-stage TVD2 and the
third-order three-stage TVD3 scheme from \citet{ShuOsher1988}. The fourth
explicit scheme is the second-order three-stage scheme from
\citet{Kraaijevanger1991}, further studied in
\citet{KetchesonMacdonaldGottlieb2009}
and \citet{KupkaHappenhoferHiguerasKoch2012}, called SSP\,RK(3,2).

The TVD2 and TVD3 (total variation diminishing) schemes were also analysed with
respect to their SSP (strong stability preserving) properties by
\citet{Kraaijevanger1991}. Their coefficients were first derived by
\citet{Heun1900} and \citet{Fehlberg1970} from a different viewpoint. They are
the explicit Runge--Kutta schemes of second order with two stages (TVD2) and of
third order with three stages (TVD3) which have the largest domain for which the
SSP property holds among all schemes of such order and such number of stages,
i.e.\ they are the optimum SSP\,RK(2,2) and SSP\,RK(3,3) schemes. The
SSP\,RK(3,2) scheme is the optimum one among all three-stage explicit
Runge--Kutta schemes with SSP property, if the approximate order is required to
be only two instead of three (see \citet{Kraaijevanger1991} for proofs of these
results). It can be implemented with the same memory consumption as TVD2
\citep{Ketcheson2008,HappenhoferKochKupka2011}. The Butcher arrays
\citep[e.g.,][]{Kraaijevanger1991,LeVeque2007} and the Shu--Osher arrays
\citep{ShuOsher1988} of all metioned schemes are given in Table~\ref{tab-butcher} 
resp.\ Table~\ref{tab-shuosher}.

We note that all schemes are explicit schemes. According to 
\citet{WangSpiteri2007}, they are all linearly unstable in theory when coupled
with the WENO5 scheme except TVD3. But the Courant numbers we use are small
enough in terms of \citet{MotamedMacdonaldRuuth2011} to make the combination
with WENO5 stable in practical applications.

\begin{table}[ht]
\centering
\begin{tabular}{c|cc} 
$0$            &               &               \\[1mm] 
$1$            & $1$           &               \\[1mm] 
\hline                                         \\[-3mm]  
$A_{\rm TVD2}$ & $\frac{1}{2}$ & $\frac{1}{2}$
\end{tabular}
\begin{tabular}{c|ccc} 
$0$                    &               &               &               \\[1mm]  
$\frac{1}{2}$          & $\frac{1}{2}$ &               &               \\[1mm]  
$1$                    & $\frac{1}{2}$ & $\frac{1}{2}$ &               \\[1mm]  
\hline                                                                 \\[-3mm]  
$A_{\rm SSP\,RK(3,2)}$ & $\frac{1}{3}$ & $\frac{1}{3}$ & $\frac{1}{3}$
\end{tabular}
\begin{tabular}{c|ccc} 
$0$            &               &               &               \\[1mm] 
$1$            & $1$           &               &               \\[1mm] 
$\frac{1}{2}$  & $\frac{1}{4}$ & $\frac{1}{4}$ &               \\[1mm] 
\hline                                                         \\[-3mm]  
$A_{\rm TVD3}$ & $\frac{1}{6}$ & $\frac{1}{6}$ & $\frac{2}{3}$
\end{tabular}
\caption{The Butcher arrays of the explicit schemes considered in this paper. 
         From left to right: TVD2, SSP\,RK(3,2), TVD3.} \label{tab-butcher}
\end{table}

\begin{table}[ht]
\centering
\begin{tabular}{ccccccccc} 
scheme & order & stages & \multicolumn{3}{c}{$\alpha_i$} & \multicolumn{3}{c}{$\beta_i$} \\[1mm] 
\hline \\[-3mm] 
Euler        & 1 & 1 & 1             &               &               & 1             &               &               \\[1mm] 
\hline \\[-3mm]
TVD2         & 2 & 2 & 1             &               &               & 1             &               &               \\[1mm] 
             &   &   & $\frac{1}{2}$ & $\frac{1}{2}$ &               & 0             & $\frac{1}{2}$ &               \\[1mm] 
\hline \\[-3mm]
SSP\,RK(3,2) & 2 & 3 & 1             &               &               & $\frac{1}{2}$ &               &               \\[1mm] 
             &   &   & 0             & 1             &               & 0             & $\frac{1}{2}$ &               \\[1mm] 
             &   &   & $\frac{1}{3}$ & 0             & $\frac{2}{3}$ & 0             & 0             & $\frac{1}{3}$ \\[1mm] 
\hline \\[-3mm]
TVD3         & 3 & 3 & 1             &               &               & 1             &               &               \\[1mm] 
             &   &   & $\frac{3}{4}$ & $\frac{1}{4}$ &               & 0             & $\frac{1}{4}$ &               \\[1mm] 
             &   &   & $\frac{1}{3}$ & 0             & $\frac{2}{3}$ & 0             & 0             & $\frac{2}{3}$
\end{tabular}
\caption{The Shu--Osher arrays \citep{ShuOsher1988} of the explicit schemes considered 
         in this paper.} 
         \label{tab-shuosher}
\end{table}

\section{Analytical Test Cases}
\label{sec-analytical}

In practice, the order of accuracy is not sufficient to describe the efficiency
of a Runge--Kutta method. As described in Appendix~A.6 in \citet{LeVeque2007}, we 
assume that the error $\varepsilon$ of a method decays with the step size $h$ as

\begin{equation}
  \varepsilon(h) \approx C h^p, \label{eq-errdecay}
\end{equation}

\noindent where $p$ is the {\it (empirical) order of convergence} or {\it order
of accuracy} and $C$ is the {\it error constant} of the method. $\varepsilon(h)$ 
is the numerical error at grid spacing $h$. A higher order method may, for a given 
grid, deliver worse results than a lower order scheme due to its high error constant 
$C$ \citep[p.\ 35,][]{FerzigerPeric2002}.

$p$ and $C$ can be estimated from a numerical solution if the exact (or at least,
very accurate) solution is known by comparing the error for several values of $h$.
If, for example, the resolution is increased by a factor $2$, the convergence rate 
$p$ can be estimated by 

\begin{equation}
  p = \log_2 \left( \varepsilon(h) / \varepsilon(h/2) \right).
\end{equation}
 
Then, the error constant of the method can be calculated by

\begin{equation}
  C = \varepsilon (h) / h^p. \label{eq-errcons}
\end{equation}

The obtained values depend on the test problem and on the norm chosen to 
measure the error size.

We compare the efficiency and accuracy of several numerical schemes 
by solving the advection equation

\begin{equation}
  \frac{\partial \phi}{\partial t} + u \frac{\partial \phi}{\partial x} = 0 
  \label{eq-adv1d}
\end{equation}

\noindent for $t \in (0, 2 ]$ and $x \in \left[ 0, 1 \right]$ with periodic 
boundary conditions. The advection velocity $u$ is set to $1$. The analytical 
solution of the advection equation~\eqref{eq-adv1d} at time $t$ is $\phi(x,t) 
= \phi( x - t, 0)$. With the initial condition

\begin{equation}
  \phi \left( x, 0 \right) = 1 + 0.1 \sin \left( 2 \pi x \right), 
  \label{eq-sinic}
\end{equation}

\noindent the analytical solution stays smooth for all times. Therefore, this is an 
appropriate test case for determining the empirical order of accuracy and the error 
constants of a method.

Given discontinuous initial data,

\begin{equation}
  \phi \left( x, 0 \right) = \begin{cases} 1, & {\rm if }\ 0.1 < x < 0.3, \\
                                           0, & {\rm else},
                             \end{cases} \label{eq-stepic}
\end{equation}

\noindent the convergence order is restricted by the smoothness of the solution. 
By comparing the numerical solution to the analytical one, we calculate the mean $L^2$ 
error at $t = 2\,{\rm s}$ \citep[cf.\ Appendix~A.5 in][]{LeVeque2007} for a set of 
spatial and temporal resolutions. 

For the advection equation, the (advective) Courant number $\sigma$ is defined by

\begin{equation}
  \sigma = \abs{u} \frac{\delta t}{\delta x}. \label{eq-couradv}
\end{equation}

Next, we solve the one-dimensional diffusion equation

\begin{equation}
  \frac{\partial \phi}{\partial t} - D \frac{\partial^2 \phi}{\partial x^2} = 0 
  \label{eq-diff1d}
\end{equation}

\noindent for $t \in (0, 50 ]$ and $x \in \left[ 0, 1 \right]$ with periodic 
boundary conditions and with initial data

\begin{equation}
  \phi \left( x, 0 \right) = 1.1 + 0.1 \sin \left( 2 \pi x \right).
  \label{diff-sinic}
\end{equation}

The analytical solution is

\begin{equation}
  \phi \left( x, t \right) = 1.1 + 0.1 \exp \left( - D \pi^2 t \right) 
                                       \sin \left( 2 \pi x \right).
\end{equation}

$D>0$ is the (constant) diffusion coefficient which we choose as $10^{-3}$. 
For the diffusion equation, we define the (diffusive) Courant number $\sigma$ by

\begin{equation}
  \sigma = D \frac{\delta t}{\delta x^2}. \label{eq-courdiff}
\end{equation}

\subsection{Errors of Runge--Kutta Schemes}
\label{sec-temporal}

For the following, we choose the fifth order WENO scheme to discretise the advection
equation in space, and compare the efficiency of several Runge--Kutta schemes for the 
analytical test problems from Section~\ref{sec-analytical}. The results are given for 
the smooth initial condition~\eqref{eq-sinic} in Tables~\ref{tab-adv1Euler}, 
\ref{tab-adv1TVD2}, \ref{tab-adv1TVD3} and~\ref{tab-adv1SSPRK32} for the Euler forward, 
the TVD2, the TVD3, and the SSP\,RK(3,2) scheme, respectively. For the discontinuous 
initial condition~\eqref{eq-stepic}, they can be found in Tables~\ref{tab-adv2Euler}, 
\ref{tab-adv2TVD2}, \ref{tab-adv2TVD3} and~\ref{tab-adv2SSPRK32}. In each row, the spatial 
resolution is fixed, whereas in the columns, the temporal resolution is constant. Since 
$\phi$ is of magnitude $1$, the absolute errors shown are also relative errors. 

For advective problems, we see the error for a fixed Courant number on the diagonal of 
each of the error tables. If the solution was not stable for the particular choice of 
spatial and temporal resolution, we do not give a number for the error size. In most 
cases, the algorithm is stable only if $\sigma < 1$. 

From these data we can deduce the size of the temporal and spatial error for each scheme, 
and its dependence on the Courant number. For the smooth initial condition~\eqref{eq-sinic},
we observe that the error $\varepsilon(\delta x,\delta t)$ of the Euler scheme is never 
smaller than about $10^{-4}$. It shows approximately first order convergence in time. For 
many combinations of $\delta t$ and $\delta x$, decreasing $\delta x$ does not lead to a 
decrease in the error, since the error is dominated by the error of the time integration 
scheme. We conclude that the Euler forward scheme is not efficient unless the spatial 
resolution is very coarse. Then, the maximum allowed Courant numbers are rather small.

For the other schemes, the error reaches much smaller magnitudes down to approximately
machine precision. In most cases, temporal and spatial error are balanced or the 
spatial error dominates except for the regions where both resolutions are either very 
coarse or very fine. The TVD3 scheme shows the smallest errors, but these are only 
reached with very high spatial and temporal resolution.

For the discontinuous problem, the errors are much larger. For a large range of combinations 
of $\delta t$ and $\delta x$, the error is nearly independent of the temporal scheme.
It does decrease with spatial resolution, but at a much slower rate determined by the 
smoothness of the solution. Nevertheless, the stability properties are different. The Euler 
forward scheme is always unstable for $\sigma \geq 1$, except for very coarse resolution,
and shows non-monotonic error convergence. All other schemes give stable solutions with 
$\sigma = 1$, but often they are very inaccurate. 

\begin{figure*}
\begin{minipage}{0.475\textwidth}
  \centering
  \includegraphics[width=1.0\columnwidth]{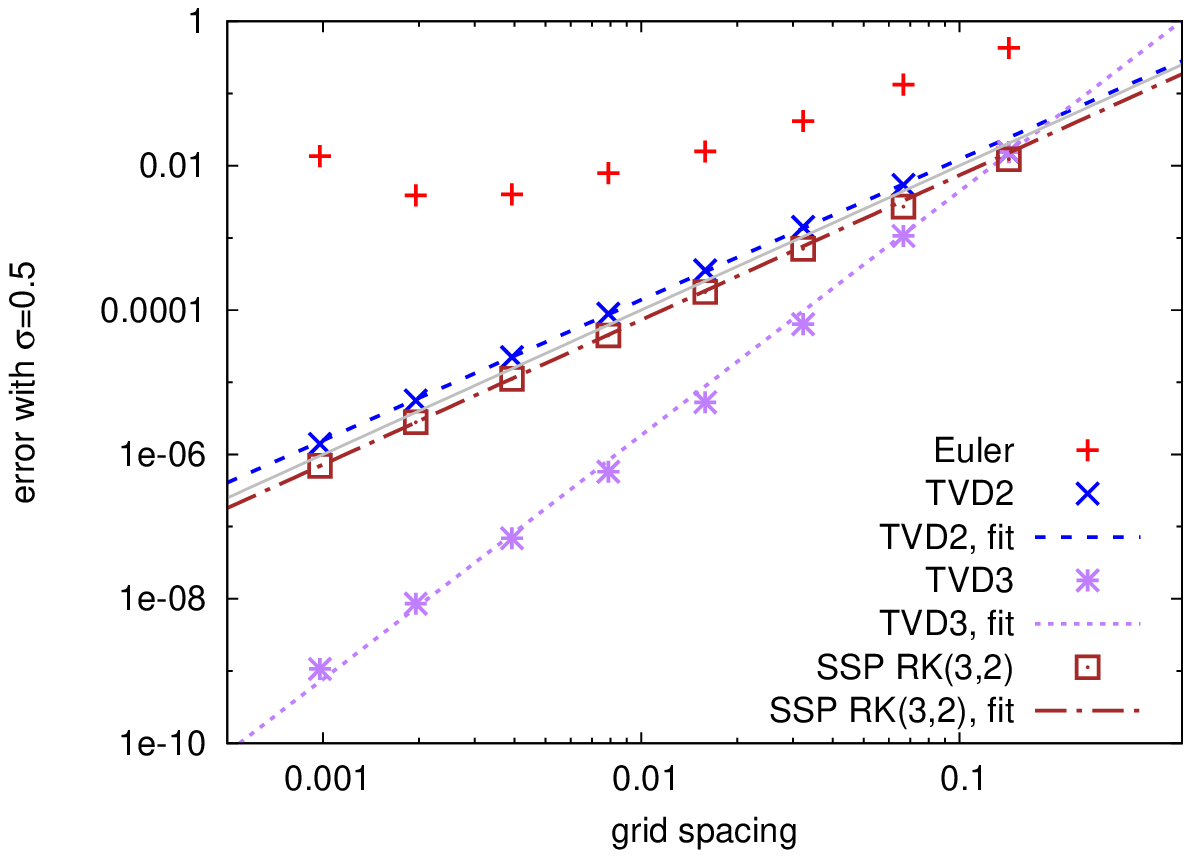}
\end{minipage}
\hfill
\begin{minipage}{0.475\textwidth}
  \centering
  \includegraphics[width=1.0\columnwidth]{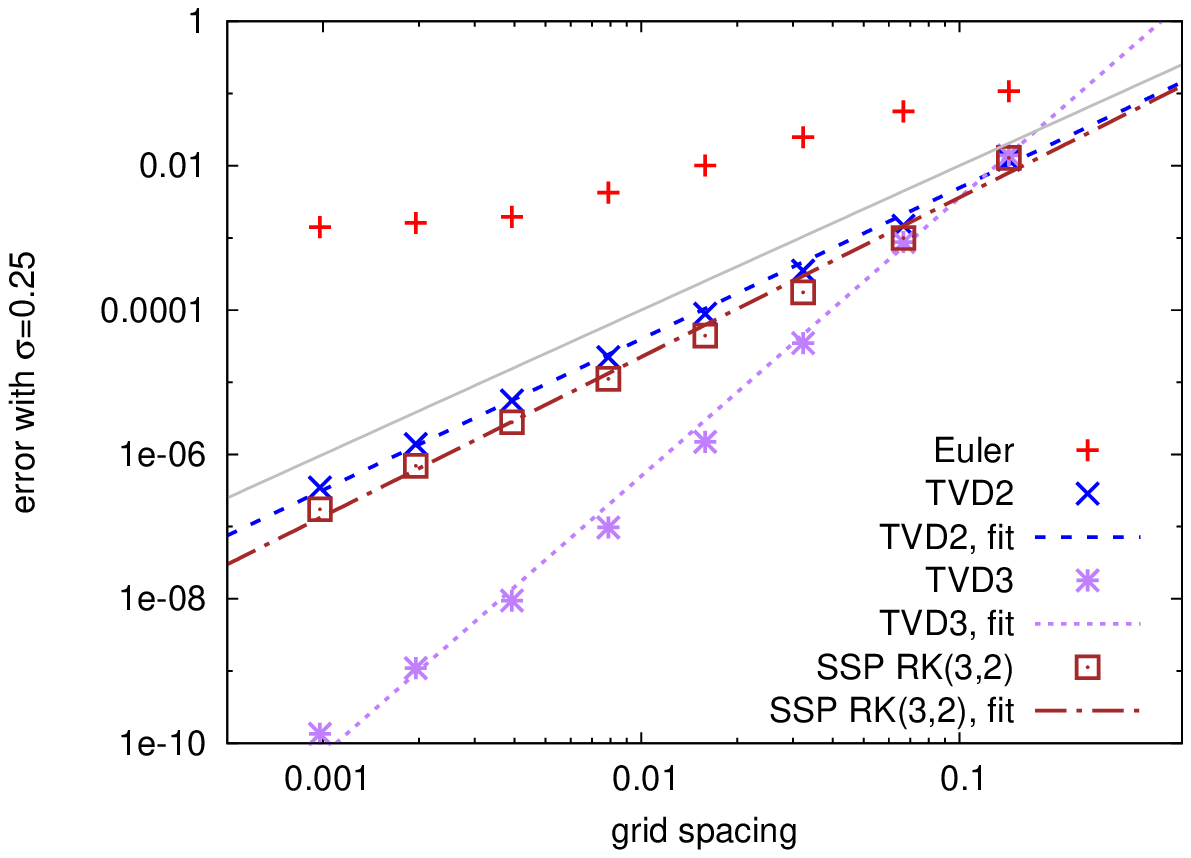}
\end{minipage}
\vskip 0.25cm
\begin{minipage}{0.475\textwidth}
  \centering
  \includegraphics[width=1.0\columnwidth]{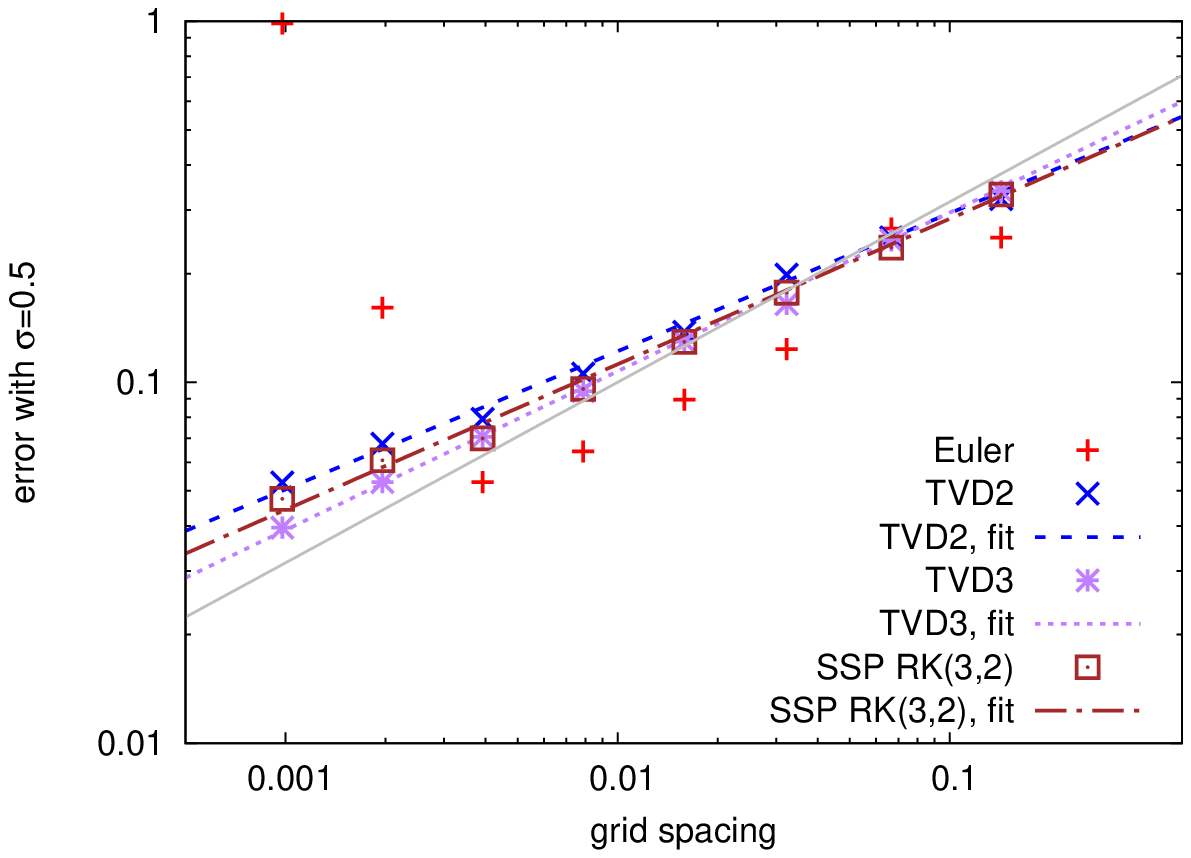}
\end{minipage}
\hfill
\begin{minipage}{0.475\textwidth}
  \centering
  \includegraphics[width=1.0\columnwidth]{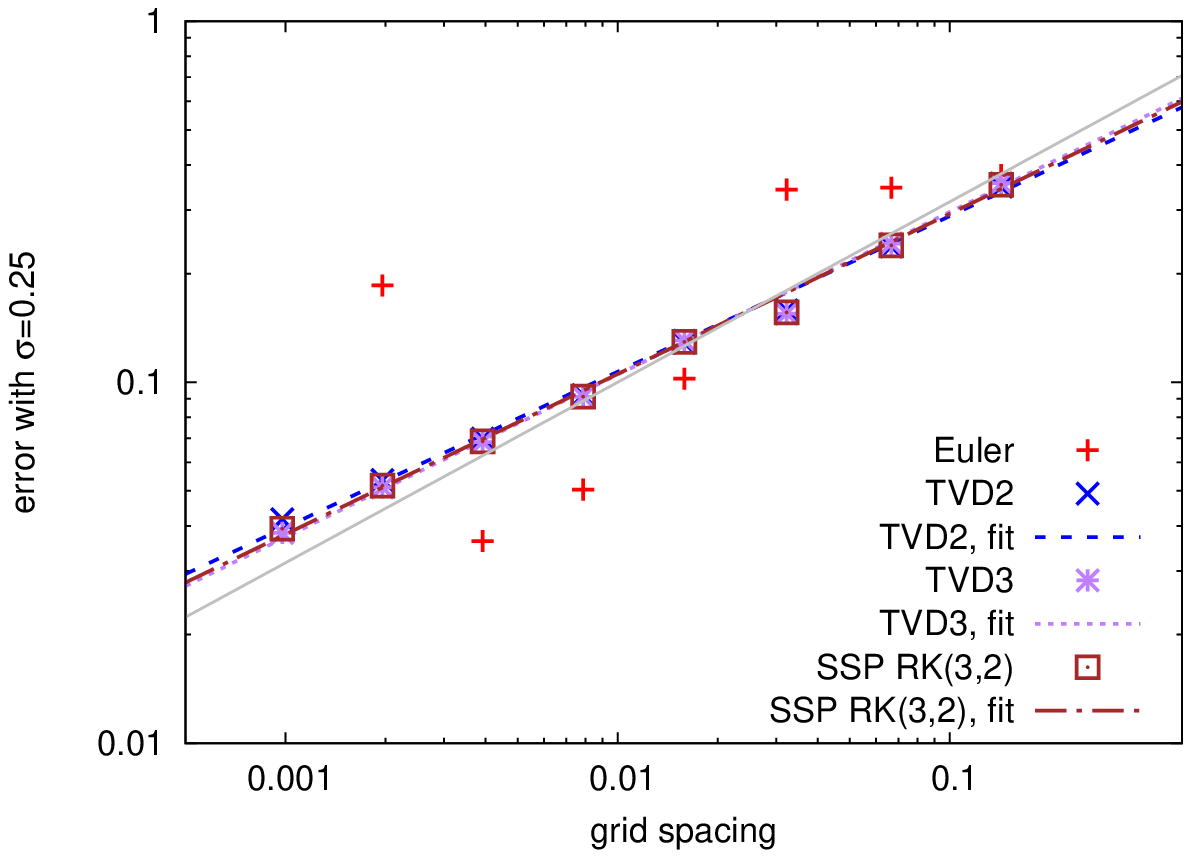}
\end{minipage}
\caption{Comparison of error convergence fits and actual error data for $\sigma
         = 0.5$ (left) and $\sigma = 0.25$ (right). Spatial discretisation
         is done with the WENO5 scheme. Top panel: advection equation with
         smooth initial data~\eqref{eq-sinic}. The grey line indicates
         second-order convergence. Bottom panel: advection equation with
         discontinuous initial data~\eqref{eq-stepic}. The grey line indicates
         square-root convergence. Euler forward is unstable for small grid
         spacing for both Courant numbers and no fit is shown for this method as
         it would not apply to the whole range of grid spacings.}
\label{fig-advfit}
\end{figure*}

In Figure~\ref{fig-advfit}, we show the error convergence of the numerical 
solution of the advection equation with smooth initial data~\eqref{eq-sinic} (top 
panel) and with discontinuous initial data~\eqref{eq-stepic} (middle panel) for 
the four Runge--Kutta scheme considered in this paper. We also show fits of the 
form~\eqref{eq-errdecay} to the error convergence. The parameters $C$ and $p$ 
of the fits are given in Table~\ref{tab-adv1errorsfixC} for the smooth case and
in Table~\ref{tab-adv2errorsfixC} for the discontinuous case. For a given set of 
pairs $(h_i, \varepsilon_i)$, the fitting parameters $C$ and $p$ as defined in 
equation~\eqref{eq-errdecay} are obtained by solving the linear system

\begin{equation}
    \left( \begin{array}{cc}
             \log_{10} h_1 & 1 \\
             \log_{10} h_2 & 1 \\
             \vdots & \vdots \\
             \log_{10} h_n & 1
           \end{array} \right) 
    \left( \begin{array}{c}
              p \\
              \log_{10} C
           \end{array} \right)
  = \left( \begin{array}{c}
             \log_{10} \varepsilon_1 \\
             \log_{10} \varepsilon_2 \\
             \vdots \\
             \log_{10} \varepsilon_n
           \end{array} \right).
\end{equation}

In the smooth case, WENO5 with Euler forward time integration yields first order 
convergence, whereas the combination with TVD2 and SSP\,RK(3,2) converges with 
second order. On average, the error constant of SSP\,RK(3,2) is smaller than 
the one of TVD2. Using TVD3 as time integrator results in third order convergence. 

For the discontinuous initial condition~\eqref{eq-stepic} the convergence order and 
the error constant for all integration schemes are very similar except for the Euler 
method. The accuracy of the numerical solution is limited by the smoothness of the
analytical solution. Only the Euler forward method shows different convergence rates 
with much strongly varying error constants, indicating very irregular error convergence. 
The additional effort of using a three-stage scheme does not pay off in terms of 
accuracy compared to TVD2. In terms of stability, TVD2, TVD3 and SSP\,RK(3,2) are 
quite similar.

From Figure~\ref{fig-advfit}, we investigate the influence of the Courant 
number $\sigma$ on the accuracy of the numerical solution. On the left panels,
$\sigma = 0.5$, whereas on the right panels, $\sigma = 0.25$. In the smooth
case, halving $\sigma$ leads to a decrease in error size by a factor $4$ for the
two second-order schemes and a factor $8$ for TVD3, whereas the changes are much
smaller for the discontinuous problem. We conclude that the accuracy of the
numerical solution is in the smooth case limited by the time integration scheme
for fine grid spacing and by the spatial discretisation for coarse grid spacing,
but in the discontinuous problem by the spatial accuracy only (for those
schemes which are actually stable).

\begin{table*}[ht]
\begin{minipage}{1.0\textwidth}
\centering
\begin{tabular}{l|ll|ll|ll|ll|ll}
             & \multicolumn{2}{c|}{$\sigma=1.0$} & \multicolumn{2}{c|}{$\sigma=0.5$} 
             & \multicolumn{2}{c|}{$\sigma=0.25$} 
             & \multicolumn{2}{c|}{$\sigma=0.125$} & \multicolumn{2}{c}{$\sigma=0.0625$} \\
scheme       & $p$ & $C$ & $p$ & $C$ & $p$ & $C$ & $p$ & $C$ & $p$ & $C$ \\
\hline 
Euler        &      &         &      &         &      &         & 0.98 & 3.10e-1 & 0.83 & 7.88e-2 \\ 
TVD2         & 0.90 & 2.97e-1 & 1.95 &  1.08e0 & 2.09 & 6.09e-1 & 2.34 & 6.35e-1 & 2.66 & 9.69e-1 \\ 
TVD3         & 3.05 &  1.24e1 & 3.38 &  1.05e1 & 3.86 &  2.57e1 & 4.35 &  9.82e1 & 4.68 &  2.76e2 \\ 
SSP\,RK(3,2) & 1.11 & 1.62e-1 & 2.00 & 7.42e-1 & 2.21 & 5.84e-1 & 2.49 & 7.53e-1 & 2.83 &  1.36e0
\end{tabular}
\caption{Empirical order of accuracy $p$ and error constants $C$ for WENO with 
         several time integration schemes and fixed Courant numbers $\sigma$ when 
         solving~\eqref{eq-adv1d} \& \eqref{eq-sinic}.}
\label{tab-adv1errorsfixC}
\end{minipage}
\end{table*}

\begin{table*}[ht]
\begin{minipage}{1.0\textwidth}
\centering
\begin{tabular}{l|ll|ll|ll|ll|ll}
             & \multicolumn{2}{c|}{$\sigma=1.0$} & \multicolumn{2}{c|}{$\sigma=0.5$} 
             & \multicolumn{2}{c|}{$\sigma=0.25$} 
             & \multicolumn{2}{c|}{$\sigma=0.125$} & \multicolumn{2}{c}{$\sigma=0.0625$} \\
scheme       & $p$ & $C$ & $p$ & $C$ & $p$ & $C$ & $p$ & $C$ & $p$ & $C$ \\
\hline
Euler        &      &         &       &         &      &         & 0.60 &  1.31e0 & 0.56 &  1.05e0 \\ 
TVD2         & 0.30 & 8.63e-1 &  0.38 & 7.09e-1 & 0.43 & 7.79e-1 & 0.45 & 8.23e-1 & 0.45 & 8.23e-1 \\ 
TVD3         & 0.27 & 6.41e-1 &  0.44 & 8.12e-1 & 0.45 & 8.35e-1 & 0.45 & 8.30e-1 & 0.45 & 8.24e-1 \\ 
SSP\,RK(3,2) & 0.37 & 7.51e-1 &  0.40 & 7.18e-1 & 0.44 & 8.14e-1 & 0.45 & 8.27e-1 & 0.45 & 8.24e-1 \\ 
\end{tabular}
\caption{Empirical order of accuracy $p$ and error constants $C$ for WENO with 
         several time integration schemes and fixed Courant numbers $\sigma$ when 
         solving~\eqref{eq-adv1d} \& \eqref{eq-stepic}.}
\label{tab-adv2errorsfixC}
\end{minipage}
\end{table*}

\begin{figure*}
\begin{minipage}{0.475\textwidth}
  \centering
  \includegraphics[width=1.0\columnwidth]{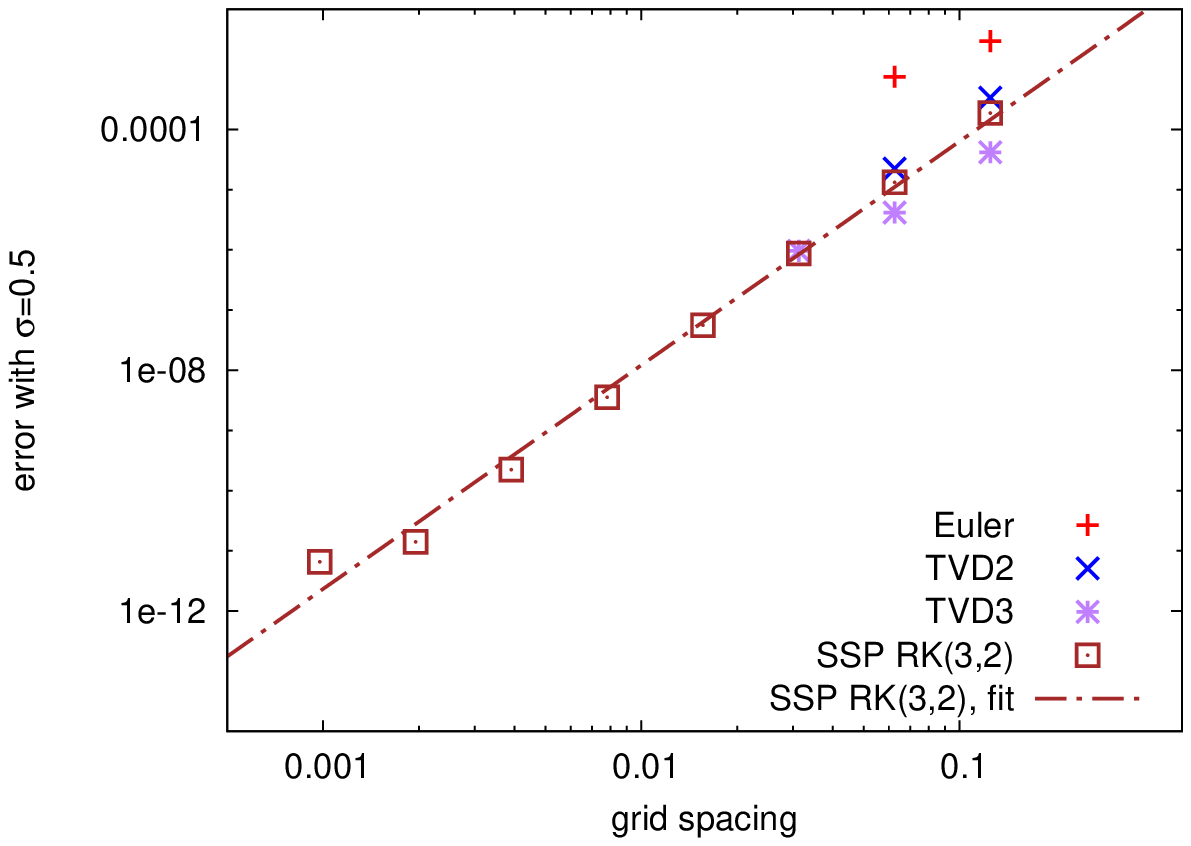}
\end{minipage}
\hfill
\begin{minipage}{0.475\textwidth}
  \centering
  \includegraphics[width=1.0\columnwidth]{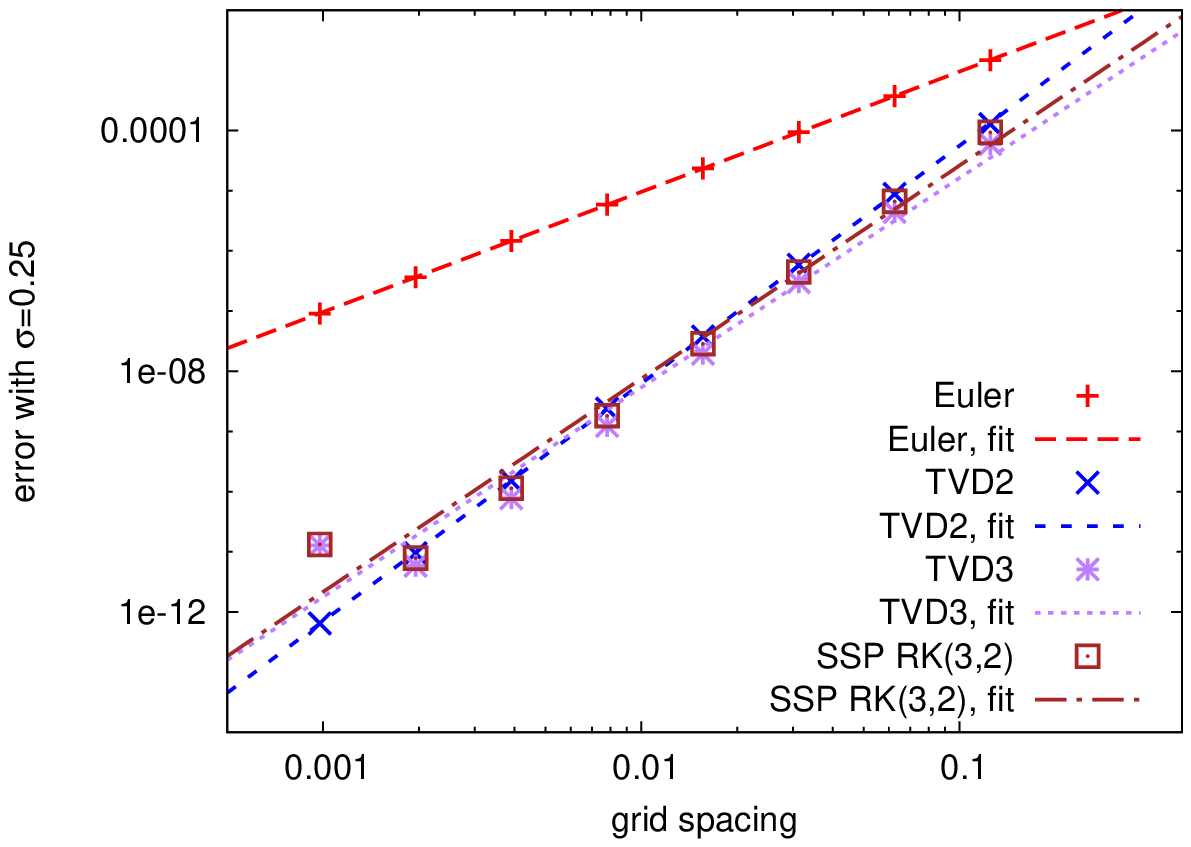}
\end{minipage}
\caption{Comparison of error convergence fits and actual error data for $\sigma = 0.5$ (left) 
         and $\sigma = 0.25$ (right) for the diffusion equation with smooth initial
         data~\eqref{eq-sinic}.}
\label{fig-difffit}
\end{figure*}

For the diffusion test case~\eqref{eq-diff1d} with the smooth initial condition~\eqref{diff-sinic},
the errors of the numerical solutions calculated at $t = 50\, {\rm s}$ are shown in 
Tables~\ref{tab-diffEuler}, \ref{tab-diffTVD2}, \ref{tab-diffTVD3} and~\ref{tab-diffSSPRK32}.
The error plots for fixed Courant numbers $\sigma$ are shown in Figure~\ref{fig-difffit}
together with fits of the form~\eqref{eq-errdecay}. The parameters $C$ and $p$ of the 
fits are given in Table~\ref{tab-differrorsfixC}.

Keeping $\sigma$ fixed and decreasing the grid spacing means decreasing the time step 
size quadratically. Therefore, the empirical convergence rates and error constants 
shown in Table~\ref{tab-differrorsfixC} exhibit second to fourth order convergence 
since the convergence rate of the time integration is doubled. The overall convergence 
is then restricted by the fourth order spatial discretisation as defined in 
equations~\eqref{eq-diffstencils}. The increase in error for very small grid spacings 
is due to rounding errors, and is larger the more stages the time integration scheme has. 

From Table~10 in \citet{KupkaHappenhoferHiguerasKoch2012} we deduce that the maximum
Courant number $\sigma$ as defined in equation~\eqref{eq-courdiff} for diffusive terms is
$0.375$ for TVD2, $0.299$ for TVD3 and $0.672$ for SSP\,RK(3,2). This is confirmed by 
the stability behaviour of the numerical solution of the test problem~\eqref{eq-diff1d} 
with initial condition~\eqref{diff-sinic}. Only SSP\,RK(3,2) yields stable results with 
$\sigma=0.5$. We conclude that the high maximum Courant number of SSP\,RK(3,2) makes it 
the most efficient scheme for diffusion--type equations even though its theoretical order 
of accuracy in time is only $2$. We note that even with non-smooth initial data, the 
solution of the diffusion equation~ \eqref{eq-diff1d} is smooth for $t>0$ and the error 
sizes converge in the same manner \citep{Evans2002}.

One could argue that formally, it is inconsistent to measure convergence orders by using 
the spatial resolution $\delta x$ as $h$ in formula~\eqref{eq-errdecay} since the number 
of degrees of freedom increases quadratically for the advection equation or even cubic 
for the diffusion equation due to the smaller time steps induced by the CFL condition. 
But from a practical point of view, modifying the spatial resolution and choosing a Courant 
number is the only way to control the accuracy of an existing simulation. Therefore, 
measuring the order of error decay when decreasing the spatial resolution while keeping 
the Courant number fixed gives the type of ``convergence order'' which is encountered 
in applications.

\begin{table*}[ht]
\begin{minipage}{1.0\textwidth}
\centering
\begin{tabular}{l|ll|ll|ll|ll}
             & \multicolumn{2}{c|}{$\sigma=0.5$}   & \multicolumn{2}{c|}{$\sigma=0.25$} 
             & \multicolumn{2}{c|}{$\sigma=0.125$} & \multicolumn{2}{c}{$\sigma=0.0625$} \\
scheme       & $p$ & $C$ & $p$ & $C$ & $p$ & $C$ & $p$ & $C$ \\
\hline
Euler        &      &         & 2.00 & 9.49e-2 & 1.99 & 4.53e-2 & 1.98 & 2.08e-2 \\ 
TVD2         &      &         & 3.95 & 5.00e-1 & 3.91 & 2.73e-1 & 3.91 & 2.31e-1 \\ 
TVD3         &      &         & 3.48 & 4.95e-2 & 3.38 & 3.51e-2 & 3.25 & 2.31e-2 \\ 
SSP\,RK(3,2) & 3.72 & 3.34e-1 & 3.54 & 9.12e-2 & 3.39 & 4.08e-2 & 3.24 & 2.29e-2
\end{tabular}
\caption{Empirical order of accuracy $p$ and error constants $C$ for WENO with 
         several time integration schemes and fixed Courant numbers $\sigma$ when 
         solving~\eqref{eq-diff1d} \& \eqref{diff-sinic}.}
\label{tab-differrorsfixC}
\end{minipage}
\end{table*}

We note that the efficiency of the time integration scheme depends on the expected 
smoothness and the required accuracy of the numerical solution. Therefore, in the next 
section we try to estimate the typical accuracy and smoothness of a simulation of 
solar surface convection. But first, we compare the WENO spatial discretisation with 
other standard schemes for the case of the advection equation.

\subsection{Errors of WENO Schemes compared to Standard Schemes}

In this paragraph, we compare the computational efficiency of the WENO5 scheme and 
TVD2 time integration with two standard schemes: the first order accurate 
Upwind and the second order accurate Lax--Friedrichs scheme \citep{Strikwerda1989}.

On a given (equidistant) grid, the one-dimensional conservation law 

\begin{equation}
  \pad[\phi]{t} + \pad[F(\phi)]{x} = 0
\end{equation}

\noindent with the analytical flux function $F$ is discretised in space in a 
conservative fashion by \citep{Merriman2003}

\begin{equation}
  \pad[\phi_i]{t} + \frac{H_{i+\frac{1}{2}}-H_{i-\frac{1}{2}}}{\delta x} = 0.
\end{equation}

A numerical scheme defines how the numerical flux 
$H_{i+\frac{1}{2}}$ is calculated. The procedure for the Upwind, the Lax--Friedrichs 
and WENO5 scheme is summarised in Algorithms~\ref{alg-upwind}, 
\ref{alg-LxF} and~\ref{alg-weno}, respectively.

\begin{algorithm}[ht]
\caption{Calculation of the numerical flux $H_{i+\frac{1}{2}}$ with the first order
         Upwind method assuming $\pad[F]{\phi} > 0$.}
         \label{alg-upwind}
\begin{algorithmic}[1]
  \STATE $H_{i+\frac{1}{2}} = F \left( \phi_i \right)$
\end{algorithmic}
\end{algorithm}

\begin{algorithm}[ht]
\caption{Calculation of the numerical flux $H_{i+\frac{1}{2}}$ with the second order
         Lax--Friedrichs method assuming $\pad[F]{\phi} > 0$.}
         \label{alg-LxF}
\begin{algorithmic}[1]
  \STATE $H_{i+\frac{1}{2}} = 0.5 \cdot \left( \left( F(\phi_{i+1}) + F(\phi_{i}) \right)
             + \frac{\delta x}{\delta t} \cdot \left( \phi_i - \phi_{i+1} \right) \right)$
\end{algorithmic}
\end{algorithm}

\begin{algorithm*}[ht]
\caption{Calculation of the numerical flux $H_{i+\frac{1}{2}}$ with WENO5 
         assuming $\pad[F]{\phi} > 0$ \citep{Shu2003}.}
         \label{alg-weno}
\begin{algorithmic}[1]
  \STATE \begin{align*}
\beta_0 & = \frac{13}{12} \cdot \left( F(\phi_{i}) - 2 \cdot F(\phi_{i+1}) + F(\phi_{i+2}) \right)^2
          + \frac{1}{4}   \cdot \left( 3 \cdot F(\phi_{i}) - 4 \cdot F(\phi_{i+1}) + F(\phi_{i+2}) \right)^2, \\
\beta_1 & = \frac{13}{12} \cdot \left( F(\phi_{i-1}) - 2 \cdot F(\phi_{i}) + F(\phi_{i+1}) \right)^2
          + \frac{1}{4}   \cdot \left( F(\phi_{i-1}) - F(\phi_{i+1}) \right)^2, \\
\beta_2 & = \frac{13}{12} \cdot \left( F(\phi_{i-2}) - 2 \cdot F(\phi_{i-1}) + F(\phi_{i}) \right)^2
          + \frac{1}{4}   \cdot \left( F(\phi_{i-2}) - 4 \cdot F(\phi_{i-1}) + 3 \cdot F(\phi_{i}) \right)^2
         \end{align*}
  \STATE \begin{equation*}
\tilde{\omega}_0 = \frac{0.3}{\left( \epsilon + \beta_0 \right)^2},\ 
\tilde{\omega}_1 = \frac{0.6}{\left( \epsilon + \beta_1 \right)^2},\ 
\tilde{\omega}_2 = \frac{0.1}{\left( \epsilon + \beta_2 \right)^2}          
         \end{equation*}
  \STATE \begin{equation*}
\omega_0 = \frac{\tilde{\omega}_0}{\tilde{\omega}_0+\tilde{\omega}_1+\tilde{\omega}_2},\ 
\omega_1 = \frac{\tilde{\omega}_1}{\tilde{\omega}_0+\tilde{\omega}_1+\tilde{\omega}_2},\ 
\omega_2 = \frac{\tilde{\omega}_2}{\tilde{\omega}_0+\tilde{\omega}_1+\tilde{\omega}_2}
         \end{equation*}
  \STATE \begin{align*}
H_{i+\frac{1}{2}} & = \omega_0 \cdot \left( \frac{1}{3} \cdot F(\phi_{i}) 
                    + \frac{5}{6} \cdot F(\phi_{i+1}) - \frac{1}{6} \cdot F(\phi_{i+2}) \right) \\
                  & + \omega_1 \cdot \left( - \frac{1}{6} \cdot F(\phi_{i-1})
                    + \frac{5}{6} \cdot F(\phi_{i})   + \frac{1}{3} \cdot F(\phi_{i+1}) \right) \\
                  & + \omega_2 \cdot \left( \frac{1}{3} \cdot F(\phi_{i-2})
                    - \frac{7}{6} \cdot F(\phi_{i-1}) + \frac{11}{6} \cdot F(\phi_{i})  \right)
         \end{align*}
\end{algorithmic}
\end{algorithm*}

It is obvious that the complexity of the WENO scheme is much larger than for the lower 
order schemes. For calculating the numerical flux function in one grid point, only 
one evaluation of the analytical flux function $F$ is necessary for the upwind method. 
For the Lax--Friedrichs method, $2$ evaluations of $F$, $3$ additions and $2$ multiplications
are necessary. But for the WENO method, at least $5$ evaluations of $F$, $25$ additions,
$28$ multiplications, $4$ divisions and $9$ exponentiations are required (the latter can 
be expressed through $9$ multiplications since they are just of power $2$). However, given 
sufficient memory, i.e.\ if $F(\phi_i)$ can be stored, the most expensive operation in
non-academic examples, the computation of $F(\phi_i)$ has to be done only once for each 
scheme which significantly reduces the costs of particularly WENO5 in such applications.

\begin{figure*}
\begin{minipage}{0.475\textwidth}
  \centering
  \includegraphics[width=1.0\columnwidth]{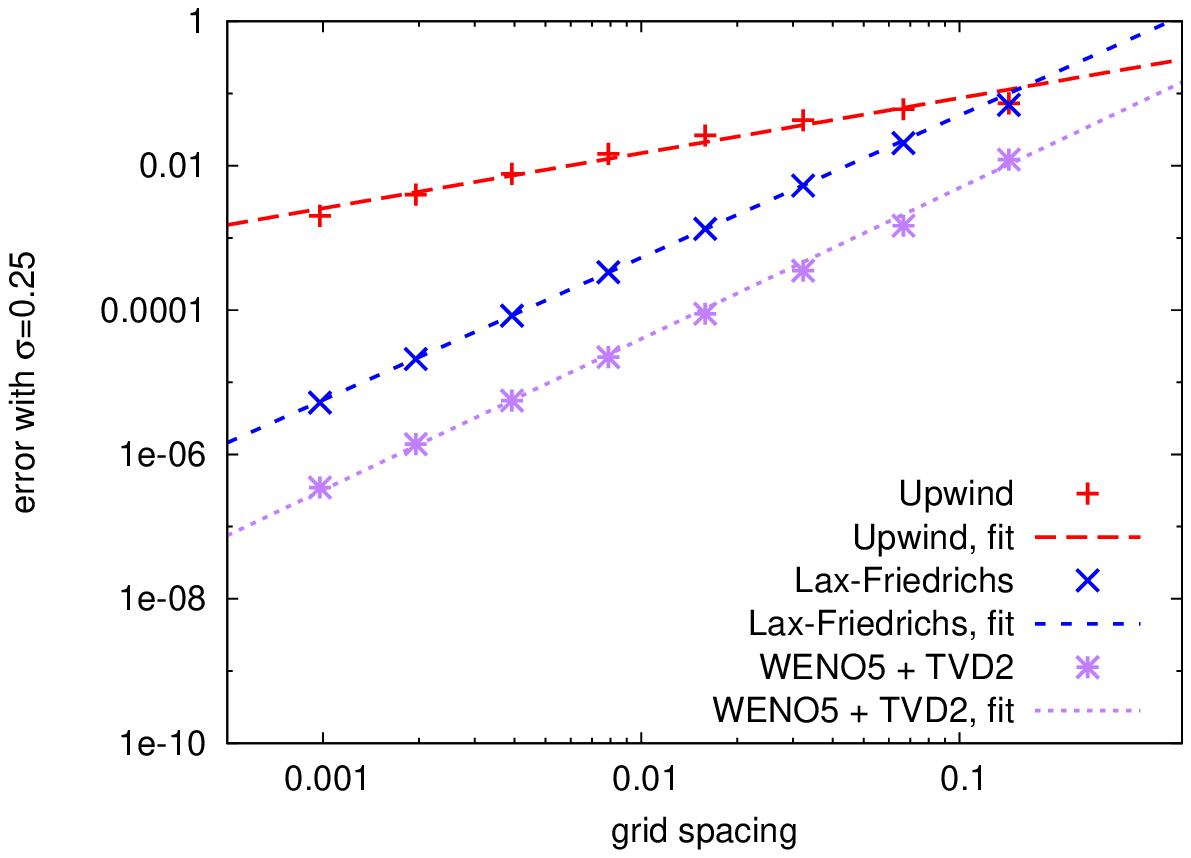}
\end{minipage}
\hfill
\begin{minipage}{0.475\textwidth}
  \centering
  \includegraphics[width=1.0\columnwidth]{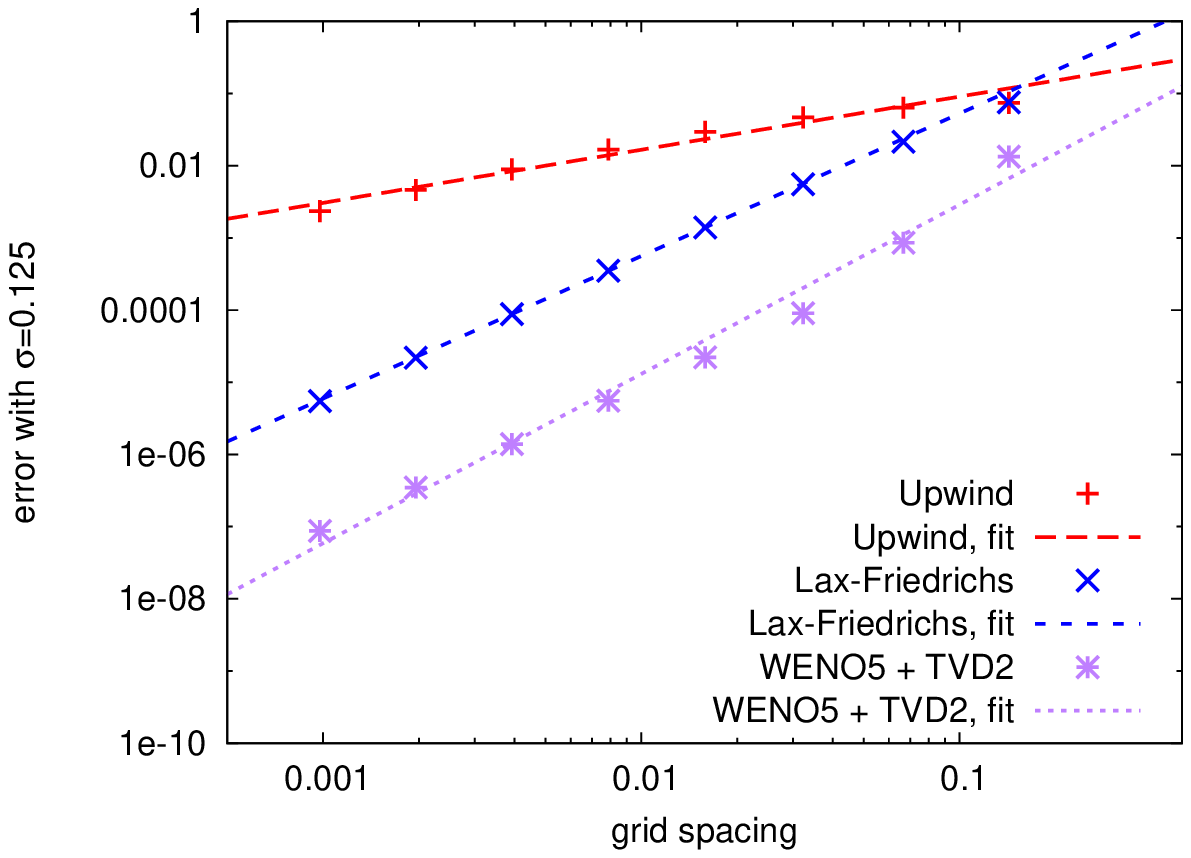}
\end{minipage}
\vskip 0.25cm
\begin{minipage}{0.475\textwidth}
  \centering
  \includegraphics[width=1.0\columnwidth]{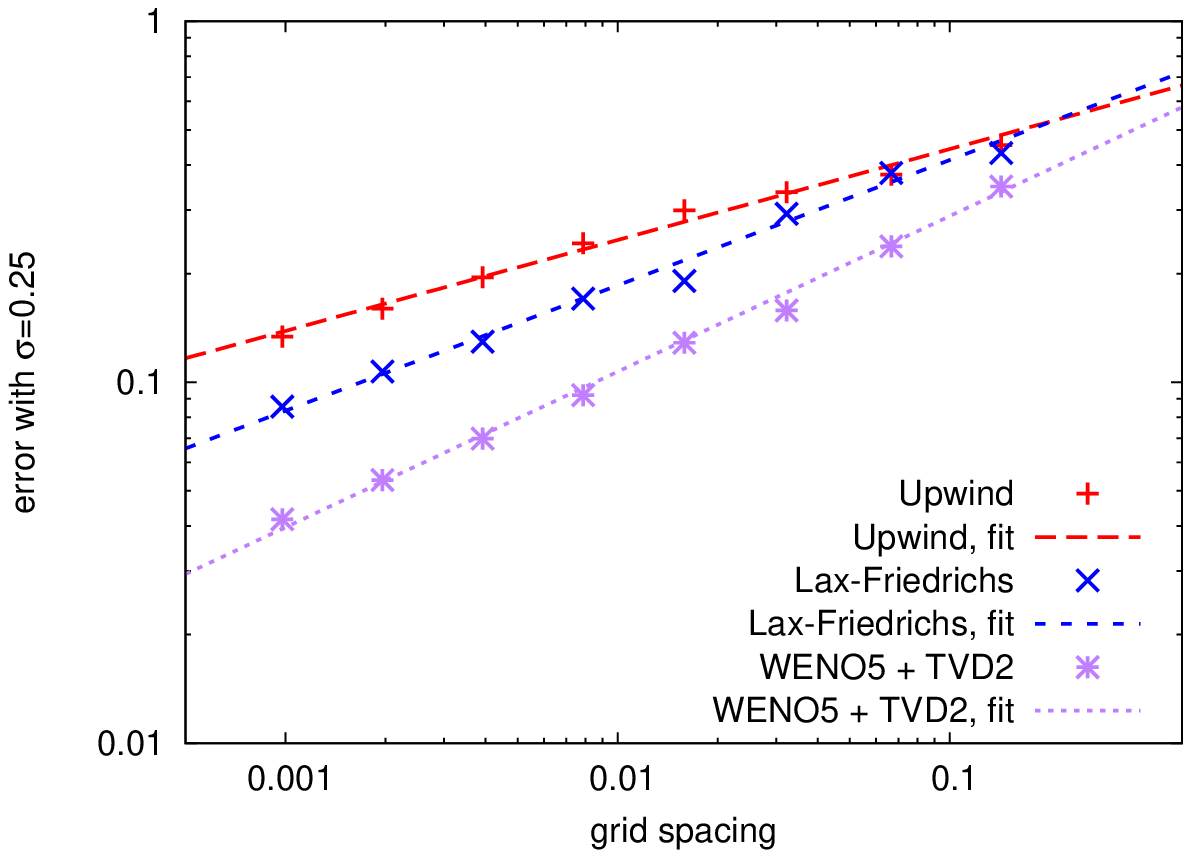}
\end{minipage}
\hfill
\begin{minipage}{0.475\textwidth}
  \centering
  \includegraphics[width=1.0\columnwidth]{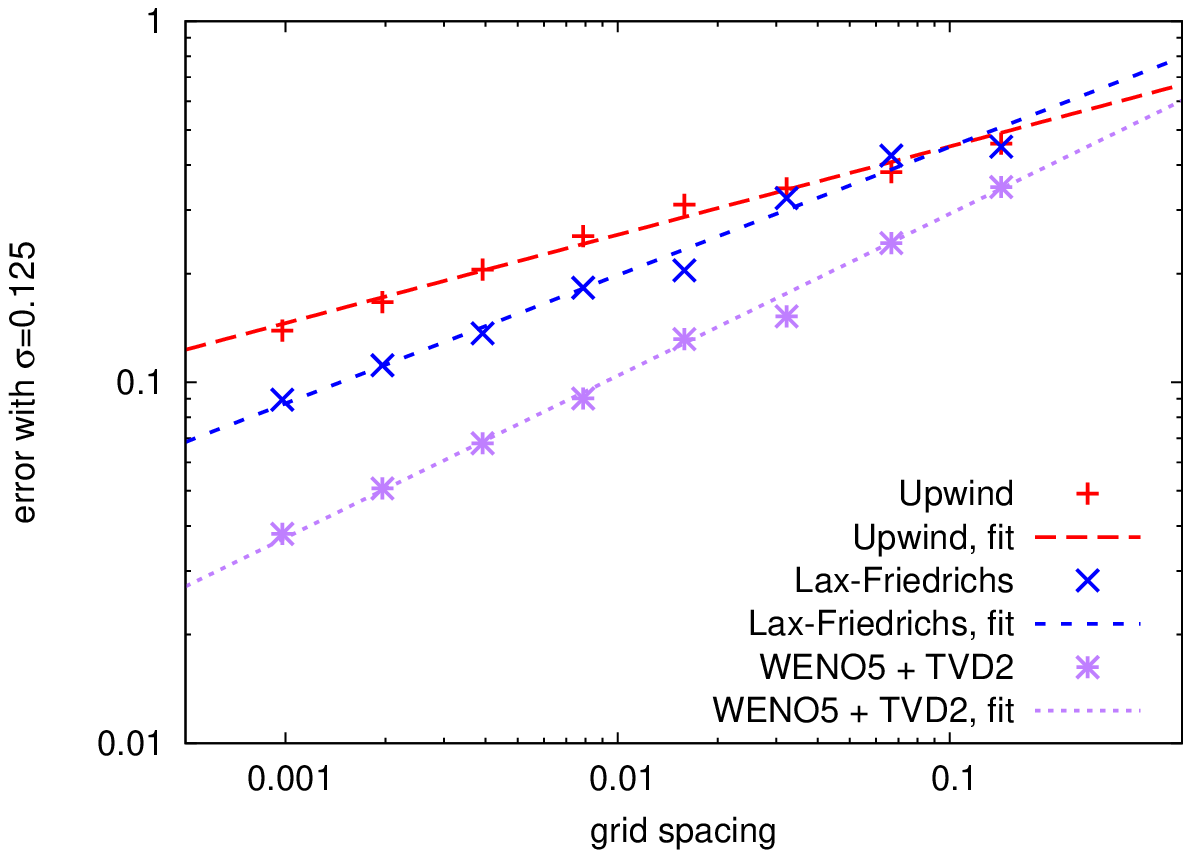}
\end{minipage}
\caption{Comparison of error convergence fits and actual error data for $\sigma = 0.25$ 
         (left) and $\sigma = 0.125$ (right) with the Upwind, the Lax--Friedrichs and the 
         WENO5 scheme with TVD2 time integration. Top panel: advection equation with smooth 
         initial data~\eqref{eq-sinic}; bottom panel: advection equation with discontinuous 
         initial data~\eqref{eq-stepic}.}
\label{fig-standfit}
\end{figure*}

In Figure~\ref{fig-standfit}, the error convergence and fits of the form~\eqref{eq-errdecay}
of the numerical solution of the advection equation~\eqref{eq-adv1d} with smooth initial 
data~\eqref{eq-sinic} and non-smooth initial data~\eqref{eq-stepic} obtained with the Upwind, 
the Lax--Friedrichs, and the WENO5 scheme combined with TVD2 time integration are shown.
The parameters $C$ and $p$ of the fits are given in Table~\ref{tab-adv1errorsfixCstandard}
for the smooth case and Table~\ref{tab-adv2errorsfixCstandard} for the discontinuous case.

From Figure~\ref{fig-standfit} we deduce that the empirical order of accuracy of the WENO5 
algorithm together with a second order time integration such as TVD2 \citep{ShuOsher1988} 
is two which is also obtained with the Lax--Friedrichs method (cf.\ \citet{Strikwerda1989}). 
Nevertheless, the error constant is much smaller than with the Lax--Friedrichs method.
In the smooth case, we observe about a factor of $8$ for $\sigma = 0.25$, whereas for 
$\sigma = 0.125$, it is about a factor of $12$. Similar is true for the discontinuous test case. 
We conclude that the numerical error obtained with WENO5 and TVD2 can be controlled
by adapting $\sigma$, whereas with the Lax--Friedrichs method, it is nearly independent
of $\sigma$.
The error of any first order method as, e.g., the upwind method (cf.\ again 
\citet{Strikwerda1989}), is larger by several magnitudes. A very high amount of grid points 
is required for them to reach an acceptable error size.

\begin{table*}[ht]
\begin{minipage}{1.0\textwidth}
\centering
\begin{tabular}{l|ll|ll|ll|ll}
             & \multicolumn{2}{c|}{$\sigma=0.5$} & \multicolumn{2}{c|}{$\sigma=0.25$} 
             & \multicolumn{2}{c|}{$\sigma=0.125$} & \multicolumn{2}{c}{$\sigma=0.0625$} \\
scheme       & $p$ & $C$ & $p$ & $C$ & $p$ & $C$ & $p$ & $C$ \\
\hline
Upwind          & 0.83 & 4.95e-1 & 0.76 & 5.01e-1 & 0.74 & 4.92e-1 & 0.72 & 4.86e-1 \\ 
Lax--Friedrichs & 1.97 &  3.69e0 & 1.97 &  4.62e0 & 1.97 &  4.97e0 & 1.98 &  5.07e0
\end{tabular}
\caption{Empirical order of accuracy $p$ and error constants $C$ for the Upwind and the 
         Lax--Friedrichs scheme with fixed Courant numbers when solving~\eqref{eq-adv1d} 
         \& \eqref{eq-sinic}.}
\label{tab-adv1errorsfixCstandard}
\end{minipage}
\end{table*}

\begin{table*}[ht]
\begin{minipage}{1.0\textwidth}
\centering
\begin{tabular}{l|ll|ll|ll|ll}
             & \multicolumn{2}{c|}{$\sigma=0.5$} & \multicolumn{2}{c|}{$\sigma=0.25$} 
             & \multicolumn{2}{c|}{$\sigma=0.125$} & \multicolumn{2}{c}{$\sigma=0.0625$} \\
scheme       & $p$ & $C$ & $p$ & $C$ & $p$ & $C$ & $p$ & $C$ \\
\hline
Upwind          & 0.24 & 6.81e-1 & 0.25 & 7.90e-1 & 0.24 & 7.91e-1 & 0.24 & 7.86e-1 \\ 
Lax--Friedrichs & 0.33 & 7.79e-1 & 0.35 & 9.18e-1 & 0.36 &  1.02e0 & 0.37 &  1.11e0
\end{tabular}
\caption{Empirical order of accuracy $p$ and error constants $C$ for the Upwind and the 
         Lax--Friedrichs scheme with fixed Courant numbers when solving~\eqref{eq-adv1d} 
         \& \eqref{eq-stepic}.}
\label{tab-adv2errorsfixCstandard}
\end{minipage}
\end{table*}

\begin{figure*}
\begin{minipage}{0.475\textwidth}
  \centering
  \includegraphics[width=1.0\columnwidth]{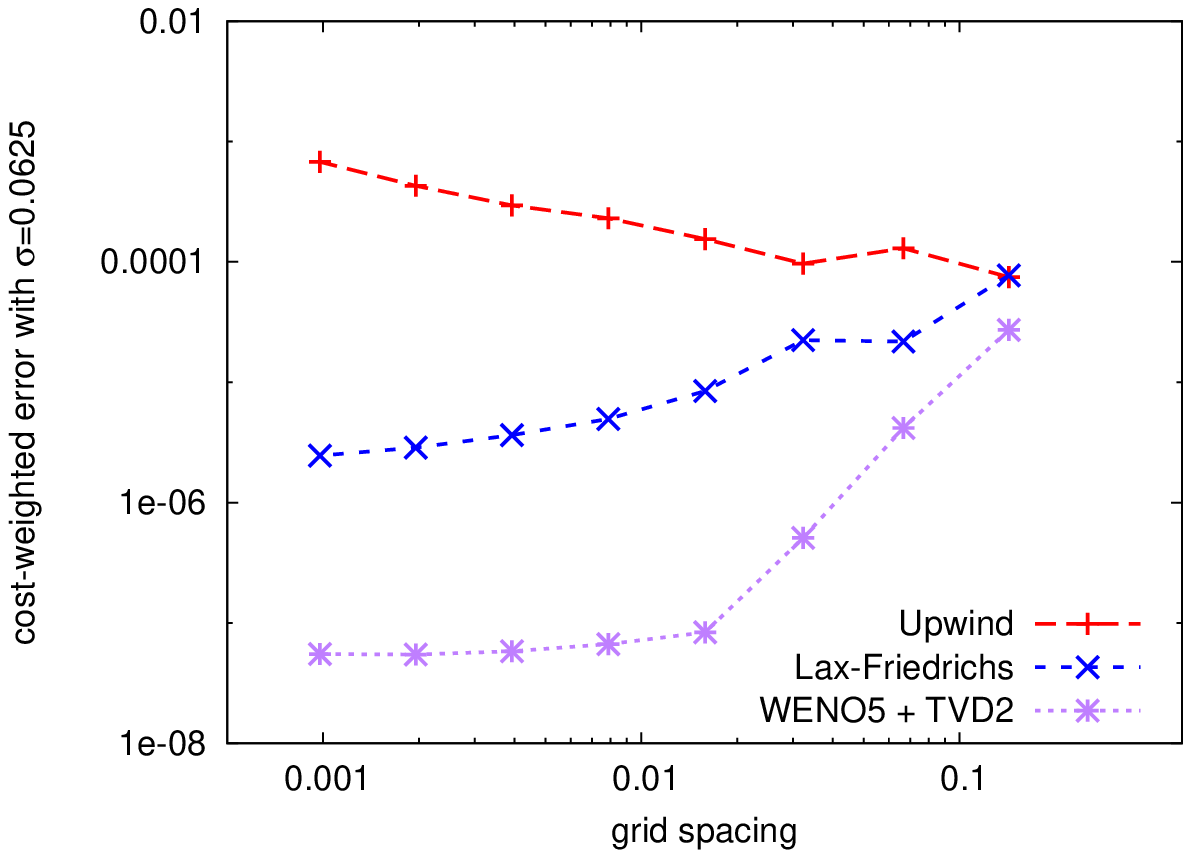}
\end{minipage}
\hfill
\begin{minipage}{0.475\textwidth}
  \centering
  \includegraphics[width=1.0\columnwidth]{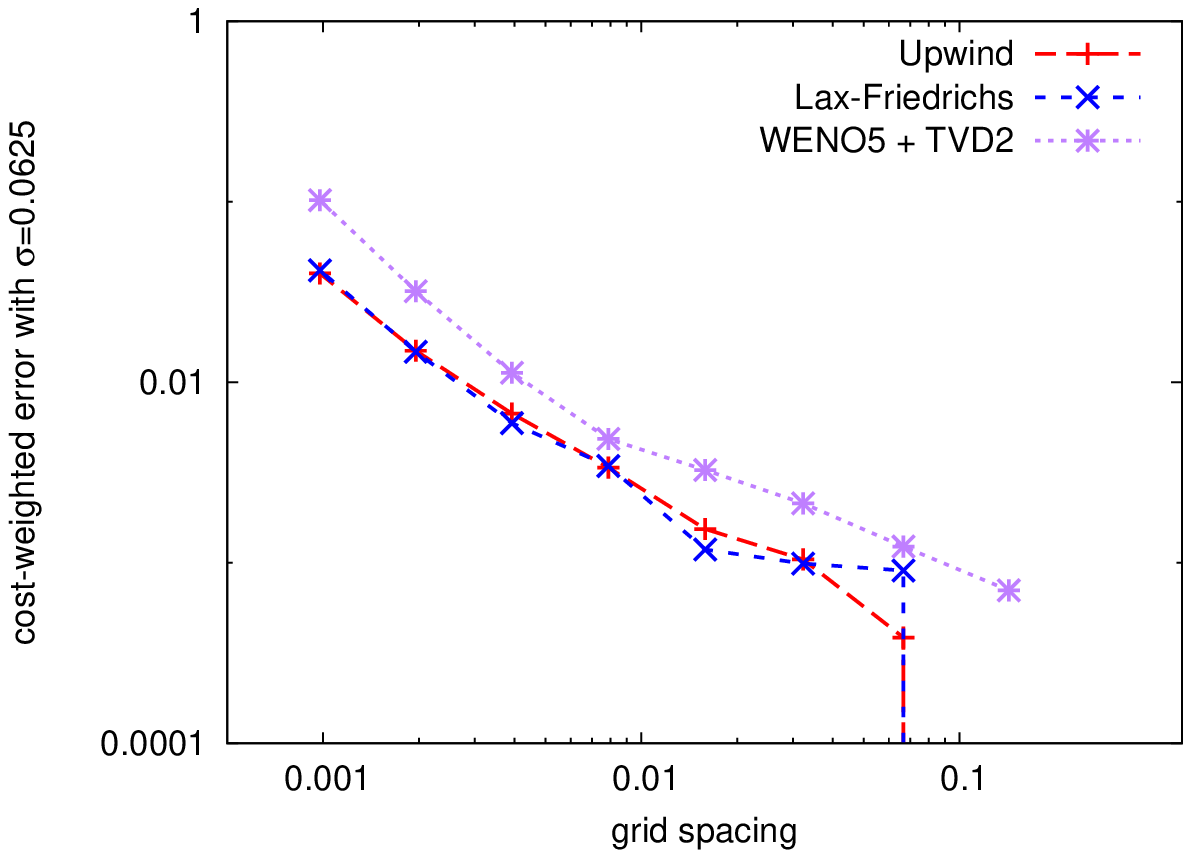}
\end{minipage}
\caption{Cost-weighted error of the numerical solution of the advection equation~\eqref{eq-adv1d}
         with smooth initial condition~\eqref{eq-sinic} and discontinuous initial 
         condition~\eqref{eq-stepic} obtained with the Upwind, the Lax--Friedrichs and the WENO5 
         scheme combined with TVD2 time integartion. Here, $\sigma = 0.0625$. The execution time
         of each simulation is measured with the \texttt{time} command and multiplied with the 
         absolute error size. For the discontinuous problem, the cost-weighted error of WENO5 
         with TVD2 is slighly higher since the increase in computation time is not compensated
         by the decrease in error. For the smooth case, however, the cost-weighted error is much 
         smaller. Furthermore, WENO5 with TVD2 gives accurate results even at much higher Courant 
         numbers.}
\label{fig-costerr}
\end{figure*}

We measure the wall clock time of the simulations performed on an Intel Core2
Duo CPU with $3.0\,{\rm GHz}$ clock rate. By multiplying the error with the wall
clock time of the simulation, we can compare the efficiency of the schemes
considered in this section. For both the smooth and non-smooth case, we show the
cost-weighted errors of the Upwind, the Lax--Friedrichs and the WENO5 scheme
with TVD2 time integration in Figure~\ref{fig-costerr}. We choose $\sigma =
0.0625$ since otherwise the wall clock times are too short to be reliable. In
the smooth case, WENO5 is much more efficient over the whole range of
resolutions considered in this test. The higher computational complexity of the
scheme leads to a more than proportional increase in accuracy. For the
discontinuous problem, the computational costs are slightly higher since the
accuracy of the numerical solution is determined by the analytical smoothness of
the solution. Nevertheless, the difference is small, and the higher complexity
of WENO5 pays off in higher stability of the scheme (the difference is a factor
of $3$ to $4$ for large grid spacing whereas it ranges from $8$ to $30$
comparing the WENO5 scheme with Lax--Friedrichs and even up to $10^4$
when comparing WENO5 to the Upwind scheme). A much higher Courant number can be
used for WENO5 and TVD2 without considerably increasing the error size, making
the scheme much more efficient. For coarse grid spacing Lax--Friedrichs and the
Upwind scheme are even unstable for the discontinuous solution or at best
equally efficient as WENO5.

Another point which we did not mention so far are memory requirements: much
more grid points are needed with lower order schemes to reach the accuracy of
WENO5 which leads to a tremendous increase in memory consumption in particular
in higher dimensions. We conclude that WENO5 schemes are not only more accurate
than standard schemes, but also computationally more efficient.

From the comparisons done in \citet{MuthsamLoew-BaselliOberscheideretal2007}
and in \citet{MuthsamKupkaLoew-Basellietal2010}, it follows that WENO5 without
artificial diffusivities yields also much more accurate results than other high-order
methods considered in their work which, however, always require such stabilisations. 

When applying the WENO method to systems of conservation laws, the state variables 
must be transformed into the eigenstate which increases the computation time. On 
the other hand, methods where no transformation is needed are less accurate 
and artificial diffusivities are necessary to stabilise the solution, e.g.\ around 
shock fronts, which at the bottom line is less efficient
\citep{MuthsamLoew-BaselliOberscheideretal2007,MuthsamKupkaLoew-Basellietal2010}. 

In this section, we compared our methods to the most basic first and second
order accurate schemes. For more rigorous comparisons concerning the spatial
discretisation, we refer, e.g., to \citet{Shu2003}. In contrast, the purpose of
these tests is to give orientation concerning the magnitude and behaviour of the
spatial error whereas we focus on the error from the time integration and the
interplay with diffusion terms.

\section{Error Size in Simulations of Solar Surface Convection}
\label{sec-realistic}

To measure the typical error in simulations of solar surface convection with
ANTARES, we performed two three-dimensional simulations which only differ in the
numerical resolution. Their specifications are summarised in
Table~\ref{tab-errmods}. The purpose of this section is to give an estimate of
the typical error size of our numerical simulations in a realistic setting. This
estimate will be used in Section~\ref{sec-compcosts} to compare the efficiency
of several time integration schemes for this particular type 
of computational problem.

\begin{table}[ht]
\centering
\begin{tabular}{l|cccc}
  & resolution 
  & grid points 
  & box size 
  & binning \\
  & $\left[ {\rm km} \right]$ & & $\left[ {\rm Mm} \right]$ & \\
\hline
Model~1 & $19.5 \times 40.0^2$ & $195 \times 150^2$ & $3.8 \times 6.0^2$ & non-grey \\
Model~2 & $9.74 \times 20.0^2$ & $389 \times 300^2$ & $3.8 \times 6.0^2$ & non-grey
\end{tabular}
\caption{Basic parameters of the two three-dimensional models from 
         Section~\ref{sec-realistic}. Model~2 was started from Model~1. The 
         data were mapped to the finer grid by interpolation, and both models
         were run for $950\,{\rm s}$. Both models use the Smagorinsky
         subgrid model \citep{Smagorinsky1963} to represent motions with scales
         smaller than the grid resolution, and a Gauss-Radau rule with 18 rays
         for the angular integration in the radiative transfer solver. They
         use the open boundary conditions \texttt{BC~3b} from         
         \citet{Grimm-StreleKupkaLoew-Basellietal2015} at the bottom, the LLNL
         equation of state \citep{RogersSwensonIglesias1996}, the non-grey
         opacities from \citet{Kurucz1993CD13,Kurucz1993CD2}, the opacity data
         from \citet{IglesiasRogers1996} for the deep interior, and the
         composition from \citet{GrevesseNoels1993}. The WENO5 scheme was used
         for spatial discretisation, and SSP\,RK(3,2) for time integration.} 
         \label{tab-errmods}
\end{table}

\begin{table*}[ht]
\centering
\begin{tabular}{lll|lll|lll|lll}
 & & & \multicolumn{3}{c|}{$t=  0\,{\rm s}$} 
     & \multicolumn{3}{c|}{$t=160\,{\rm s}$}
     & \multicolumn{3}{c} {$t=950\,{\rm s}$} \\
variable  & unit & type     & $L^1$     & $L^2$     & $L^{\infty}$ 
                            & $L^1$     & $L^2$     & $L^{\infty}$
                            & $L^1$     & $L^2$     & $L^{\infty}$ \\[1mm]
\hline                                                         \\[-3mm]  
$\rho$              & ${\rm g}\,{\rm cm}^{-3}$ & absolute 
             &  1.40e-12 &   3.74e-11 &     5.13e-8 
             &   1.39e-8 &    2.77e-8 &     9.56e-7
             &   3.64e-8 &    6.14e-8 &     9.81e-7 \\[1mm]
$T$                 & ${\rm K}$                & absolute 
             &       0.0 &        0.0 &        16.3
             &      60.8 &      179.3 &      4067.6
             &     242.5 &      624.3 &      4971.6 \\[1mm]
$\rho$              &                          & relative 
             & $0.0\,\%$ &  $0.1\,\%$ &   $4.6\,\%$
             & $2.5\,\%$ &  $5.6\,\%$ &  $80.3\,\%$
             & $6.6\,\%$ & $18.1\,\%$ & $641.4\,\%$ \\[1mm]
$T$                 &                          & relative 
             & $0.0\,\%$ &  $0.0\,\%$ &   $0.1\,\%$
             & $0.7\,\%$ &  $2.0\,\%$ &  $60.1\,\%$
             & $2.7\,\%$ &  $7.6\,\%$ &  $86.2\,\%$ \\[1mm]
$\langle T \rangle$ & ${\rm K}$                & absolute 
             &       0.0 &        0.0 &         0.0
             &      14.9 &       31.2 &       127.8
             &      22.8 &       47.3 &       228.9 \\[1mm]
$\langle T \rangle$ &                          & relative 
             & $0.0\,\%$ &  $0.0\,\%$ &   $0.0\,\%$
             & $0.2\,\%$ &  $0.4\,\%$ &   $1.7\,\%$
             & $0.3\,\%$ &  $0.8\,\%$ &   $5.0\,\%$
\end{tabular}
\caption{The error sizes in density $\rho$ and temperature $T$ calculated by   
point-wisely comparing the simulations described in Table~\ref{tab-errmods}
according to the procedure from \citet{LeVeque2007} right after the
interpolation and for two snapshots taken after $160\,{\rm s}$ and $\sim
950\,{\rm s}$ of simulation time. Both models originally coincided
on the grid points of Model 1. The norms are calculated as described in the
cited reference. The relative errors are calculated by dividing the absolute
difference by the value of Model~1. $\langle \cdot \rangle$ stands for
horizontal averaging.} \label{tab-errors2d}
\end{table*}

We remark that these simulations are Large Eddy Simulations (LES), i.e., 
they do not resolve all scales of motion. All motions with length scales smaller than
the grid resolution are modelled by the Smagorinsky subgrid model and by the numerical
viscosity of the numerical scheme. Therefore, the measurement of ``the error'' is
not trivial. Changing the resolution will change the results and convergence of the 
solution with grid spacing cannot be expected due to the chaotic nature of turbulence.

As an approximate estimate of the error size, we calculate the difference between
the solutions on the finer and the coarser grid by point-wise comparison of the values
on coinciding grid points, as described in Appendix~A.6 in \citet{LeVeque2007}. The 
resulting error estimates in several variables and several norms are shown in 
Table~\ref{tab-errors2d}.

We show the error right after the interpolation, after $\sim 160\,{\rm s}$ and
after $950\,{\rm s}$ of simulation time. After the interpolation, the error
is very small, but already after $160\,{\rm s}$ is has grown considerably.
After $950\,{\rm s}$, the two models have completely diverged, and we observe
that the $L^{\infty}$ error is much larger than the $L^1$ and $L^2$ error. This
stems from the fact that near the optical surface, the motion of the fluid is
turbulent, and changes in the numerical parameters as grid resolution produces
arbitrarily large differences. Here, the pointwise changes due to increased
resolution are large compared to the rest of the simulation, and, as indicated
by the huge numbers, can have completely diverged in the course of the
simulation.

\begin{figure}[ht]
  \centering
  \includegraphics[width=1.0\columnwidth]{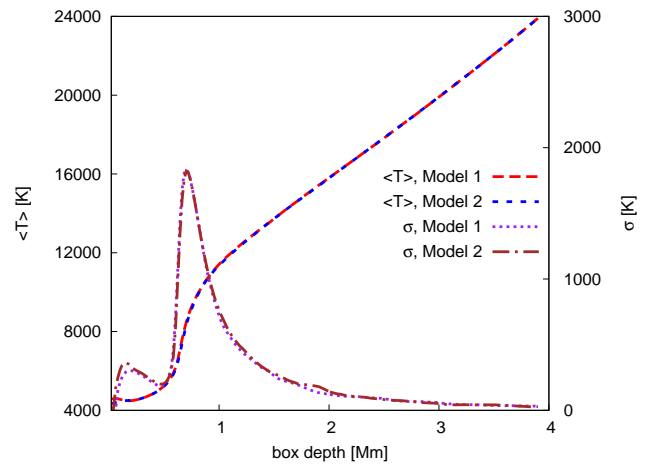}
  \caption{Mean temperature profile and standard deviation of Models~1 and~2
           as described in Table~\ref{tab-errmods}. Horizontal averages are 
           calculated after $\sim 950\,{\rm s}$ of simulation time.} 
           \label{fig-tempmom}
\end{figure}

Therefore, we refrain from measuring the error pointwisely. Instead, we suggest to use 
the mean temperature profile for error measurement. In Figure~\ref{fig-tempmom} we 
observe that the mean temperature is much more stable, but still sensitive enough 
to changes in the resolution. The standard deviation of the temperature profile 
is even more sensitive, but since it approaches $0$ it is not suited for calculating 
relative errors. Furthermore, its behaviour near the top boundary is strongly 
influenced by the boundary conditions \citep{Grimm-StreleKupkaLoew-Basellietal2015}.
The typical mean error of the temperature profile, i.e.\ the relative error in the 
$L^1$ norm, is around $0.1\,\%$ to $0.5\,\%$.

Of course, these numbers are only rough estimates. The influence
of uncertainties in the solution of the radiative transfer equation, the
equation of state, and the boundary conditions, to name only few factors, is
huge and difficult to number. Nevertheless, our tests indicate the magnitude of
the error which is far from the asymptotic regime where we can profit from the
fast error convergence of higher order time integration schemes.

In hydrodynamical simulations of similar grid size, but without the extra 
uncertainties introduced by radiative transfer and where all scales of motion
are resolved on the grid scale (i.e., DNS), the magnitude of the error
is typically of the size $0.1\,\%$ when the simulation is about to become
turbulent (cf.\ Fig.~12 to~15 in \citet{KupkaHappenhoferHiguerasKoch2012} and
Fig.~7 in \citet{HappenhoferGrimm-StreleKupkaetal2013}).

Finally, we want to determine the area ratio of smooth to non-smooth regions.
For this purpose, we calculate the nonlinearity index NI as defined in equation~(8) 
of \citet{TaylorWuMartin2007}. Therein, the nonlinear weights $\omega_j$ of the 
interpolating polynomials in the WENO reconstruction scheme as described 
in Algorithm~\ref{alg-weno} are compared to the optimal linear weights $d_j$. In 
smooth regions, they should be of the same size, whereas in non-smooth regions 
the weight of one of the parabolae should be much higher. Then, the nonlinearity 
index NI defined by

\begin{equation}
  {\rm NI} = \frac{1}{\sqrt{k(k+1)}} \left( \sum_{j=0}^k \left( 1 - 
                \frac{(k+1) \omega_j/d_j}{ \sum_{l=0}^k \omega_l/d_l} 
                                                         \right)^2 \right)^{\frac{1}{2}},
\end{equation}
 
\noindent will be close to $1$. Here, $k$ is the width of the stencil of each 
interpolation polynomial such that the order of the reconstruction process is $2k-1$.
For the fifth order variant summarised in Algorithm~\ref{alg-weno}, $k=3$.

\begin{figure}[ht]
  \centering
  \includegraphics[width=1.0\columnwidth]{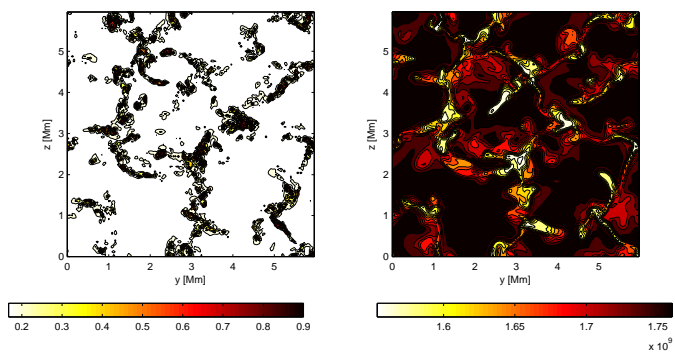}
  \caption{Left: snapshot of the nonlinearity index~NI; right: the entropy of the
           three-dimensional model, both at a geometrical depth of around $1\,{\rm Mm}$.
           The optical surface is at a geometrical depth of  around $800\,{\rm km}$.
           We observe that NI is largest in the intergranular lanes where the fluid 
           motion is most turbulent \citep{SteinNordlund2000,Kupka2009b}. On top 
           of each granule, the flow is rather smooth such that NI is small.}
           \label{fig-NIvsentropy}
\end{figure}

\begin{figure}[ht]
  \centering
  \includegraphics[width=1.\columnwidth]{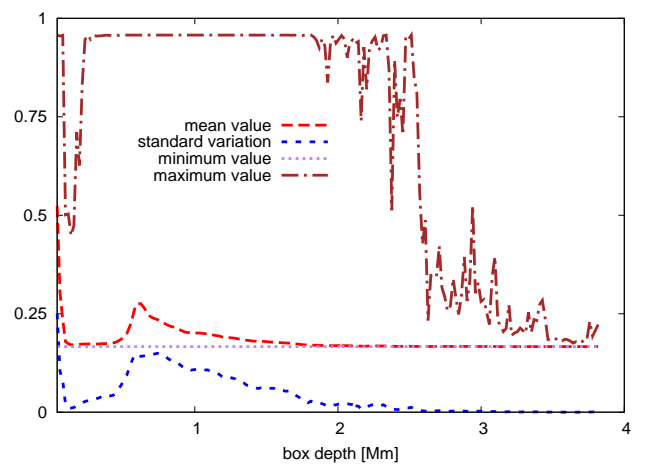}
  \caption{Mean value, standard deviation, minimum and maximum of the nonlinearity
           index NI at a specific vertical depth. We observe that NI reaches both
           its maximum values and its maximum average just below the optical surface.
           Deeper in the convection zone, the flow is smoother but NI never falls 
           below a value of around $0.17$.} \label{fig-NIstats}
\end{figure}

We plot NI as calculated in the WENO reconstruction procedure in the first
characteristic variable for reconstruction in the vertical ($x$) direction for
one snapshot of Model~1. In Figure~\ref{fig-NIvsentropy}, we show NI together
with the entropy at a fixed geometrical depth near the optical surface.
Actually, NI is located at the half-integer node, but we ignore this small
visualisation error. In Figure~\ref{fig-NIstats}, the mean value, the standard
deviation, the minimum and the maximum error in each vertical layer is plotted.

We conclude that NI captures the dynamics of surface convection very well. In
regions where the flow is turbulent --- mainly the intergranular lanes near the
optical surface (which is located at a geometrical depth of around $800\,{\rm
km}$) ---, its value is large whereas it is reasonably small in smooth regions
of the flow. We remark that the size of the minimum value of NI depends on the
design of the nonlinear weights in the WENO reconstruction
\citep{TaylorWuMartin2007}. In our tests, $\epsilon$ as defined in
Algorithm~\ref{alg-weno} is fixed to $10^{-40}$.

We conclude that even though NI is a purely numerical parameter, it also has a 
physical meaning and is a good indicator of whether a solution is smooth or not.
Counting the number of points where ${\rm NI} < 0.25$ and where ${\rm NI} > 0.5$,
we get a good estimate of the area ratio of smooth to non-smooth regions. In this 
particular simulation, the fraction of non-smooth regions never exceeds $8\,\%$ 
except for the uppermost layers which are strongly influenced by the boundary 
conditions. Over the whole simulation box, we can estimate the ratio to be

\begin{equation}
  \frac{\text{volume where the flow is non-smooth}}{\text{volume where the flow is smooth}} 
  \approx 0.05.
\end{equation}

Therefore, even though the fraction of non-smooth regions is not negligible, the 
flow in the simulation box is mostly smooth.

\section{Calculation of Computational Costs}
\label{sec-compcosts}

In this section, we want to estimate and compare the computational costs of
simulations of solar surface convection with the WENO5 scheme for spatial
discretisation and the four Runge--Kutta schemes described in
Section~\ref{sec-methods} for time integration, and under the conditions
described in Section~\ref{sec-realistic}. We leave the spatial discretisation
and the problem setup unchanged and investigate the influence of the time
integration scheme only on the accuracy and computational costs of such a
simulation.

The reason to use higher-order time integration schemes is that we expect more accurate 
results with less computation time than with the first-order Euler method. Clearly, 
all of the higher order schemes from Section~\ref{sec-methods} fulfil this for all 
grid resolutions and for both the advection and the diffusion equation.

It is more difficult to say which one of the three higher order methods is the best 
for our specific purposes. Since the time step size is determined via the 
conditions~\eqref{eq-couradv} or~\eqref{eq-courdiff} once the grid spacing is set,
the grid spacing $h$ and the Courant number $\sigma$ are the only degrees of freedom 
to control the accuracy of the numerical solution. We formulate

\begin{problem}
  Given a relative accuracy $\varepsilon_{\rm rel}$ and a Courant number $\sigma$, which 
  grid spacing is necessary for a computational cube of side length $L$, and how many 
  time steps are needed for a time interval of length $T$?
\end{problem}

Given a relative accuracy $\varepsilon_{\rm rel}$ and a Courant number $\sigma$, we can 
calculate the required relative grid spacing $h$ to reach this accuracy by interpolating 
in the tables given in~\ref{app-advsin}, \ref{app-advstp} or~\ref{app-diff}, depending 
on whether the problem is advective or diffusive and whether the solution is smooth or 
non-smooth. The grid spacing $\delta x$ of the simulation can be calculated via 
$h = \delta x / L$, where $h$ is the (relative) grid spacing obtained using the 
table interpolation. Note that $h$ is dimensionless.

In three dimensions, we need $N_h = h^{-3}$ grid points for a cube with 
side length $L$. For an advection equation with advection velocity $u$,

\begin{equation}
  N_t = n_{\rm stages}\,\frac{T}{\delta t}\ {\rm with}\ \delta t = \frac{\sigma\,\delta x}{\abs{u}} 
\end{equation}

\noindent integration steps are needed to cover a time interval of length~$T$.
The number of stages of the Runge--Kutta method is $n_{\rm stages}$, and $\sigma$ 
is the Courant number. The quantity $\alpha := \abs{u} T / L$ is a dimensionless 
quantity depending on the problem. Therefore, we define 

\begin{equation}
  N_t^{\star} := N_t / \alpha = \frac{n_{\rm stages}}{\sigma h}.
\end{equation}

Then, the computational costs $\gamma$ corresponding to the number of evaluations of 
the differential operator are given by

\begin{equation}
  \gamma = N_h \cdot N_t = N_h \cdot N_t^{\star} \cdot \alpha,
\end{equation}

\noindent or scaled by the problem-dependent factor $\alpha$,

\begin{equation}
  \gamma^{\star} = \gamma / \alpha.
\end{equation}

For a diffusion equation,
 
\begin{equation}
  N_t = n_{\rm stages}\,\frac{T}{\delta t}\ {\rm with}\ \delta t = \frac{\sigma\,\delta x^2}{D}.
\end{equation}
 
Similarly, $\beta := D\,T / L^2$ is dimensionless and problem-dependent, and we 
define the scaled number of time steps $N_t^{\star}$ by

\begin{equation}
  N_t^{\star} := N_t / \beta = \frac{n_{\rm stages}}{\sigma h^2}
\end{equation}

\noindent and the scaled computational costs $\gamma^{\star}$ by

\begin{equation}
  \gamma^{\star} = \gamma / \beta.
\end{equation}

\begin{table*}[ht]
\begin{minipage}{1.0\textwidth}
\centering
\begin{tabular}{c|rrrrrr}
scheme      & Euler & TVD2  & \multicolumn{2}{c}{TVD3} & \multicolumn{2}{c}{SSP\,RK(3,2)} \\[1mm]
$\sigma$    & 0.25  & 0.25  & 0.25  & 0.5   & 0.25  & 0.5       \\
\hline                                                          \\[-3mm]  
$\varepsilon_{\rm rel}$ & \multicolumn{6}{c}{$5 \cdot 10^{-3}$} \\[1mm]
$h$              & 8.84e-3 & 8.32e-2 & 8.24e-2 & 7.97e-2 & 8.37e-2 & 7.75e-2 \\[1mm] 
$N_h$            &  1.45e6 &  1.74e3 &  1.79e3 &  1.98e3 &  1.70e3 &  2.15e3 \\[1mm] 
$N_t^{\star}$    &  4.53e2 &  9.62e1 &  1.46e2 &  7.53e1 &  1.43e2 &  7.74e1 \\[1mm] 
$\gamma^{\star}$ &  6.56e8 &  1.67e5 &  2.61e5 &  1.49e5 &  2.44e5 &  1.66e5 
\end{tabular}
\caption{Computational costs for advective problem with smooth initial condition.}
\label{tab-compcostsadv1}
\end{minipage}
\end{table*}

In Table~\ref{tab-compcostsadv1}, the scaled computational costs with each scheme are 
calculated for an advective problem with smooth initial condition, corresponding 
to the data from~\ref{app-advsin}. According to the data from Table~\ref{tab-errors2d}, 
we choose a relative accuracy of $5 \cdot 10^{-3}$. The Euler forward scheme is by far 
the most expensive one. It needs a relative grid spacing smaller than $0.01$ to reach 
this error size. For the higher-order schemes, much larger grid spacings can be chosen. 

We observe that for TVD3 and SSP\,RK(3,2), increasing $\sigma$ from $0.25$ to $0.5$
leads to only a slightly smaller required relative grid spacing $h$. The computational 
costs are considerably decreased. Therefore, the Courant number should be set as large 
as the stability of the method allows it. Consequently, SSP\,RK(3,2) and TVD3 are the 
most efficient schemes since they allow Courant numbers of $0.5$, as indicated by the 
data in Tables~\ref{tab-adv1SSPRK32} and~\ref{tab-adv1TVD3} and confirmed by numerical 
experiments with solar surface convection simulations. TVD2 is most efficient when 
comparing all schemes with fixed Courant number of $0.25$, but it is not stable with 
higher Courant numbers. The differences in efficiency of TVD3 and SSP\,RK(3,2) are 
negligible.

With discontinuous initial data~\eqref{eq-stepic}, the error size is much larger. 
Using a relative error size of $2.5 \cdot 10^{-1}$ results in scaled computational 
costs as summarised in Table~\ref{tab-compcostsadv2}.

\begin{table*}[ht]
\begin{minipage}{1.0\textwidth}
\centering
\begin{tabular}{c|rrrrrr}
scheme      & Euler & TVD2  & \multicolumn{2}{c}{TVD3} & \multicolumn{2}{c}{SSP\,RK(3,2)} \\[1mm]
$\sigma$    & 0.25  & 0.25  & 0.25  & 0.5   & 0.25  & 0.5       \\
\hline                                                          \\[-3mm]  
$\varepsilon_{\rm rel}$ & \multicolumn{6}{c}{$2.5 \cdot 10^{-1}$} \\[1mm]
$h$              & 1.41e-2 & 6.94e-2 & 6.68e-2 & 6.44e-2 & 6.83e-2 & 7.17e-2 \\[1mm] 
$N_h$            &  3.58e5 &  2.99e3 &  3.35e3 &  3.75e3 &  3.14e3 &  2.71e3 \\[1mm] 
$N_t^{\star}$    &  2.84e2 &  1.15e2 &  1.80e2 &  9.32e1 &  1.76e2 &  8.37e1 \\[1mm] 
$\gamma^{\star}$ &  1.02e8 &  3.45e5 &  6.02e5 &  3.49e5 &  5.52e5 &  2.27e5 
\end{tabular}
\caption{Computational costs for advective problem with non-smooth initial condition.}
\label{tab-compcostsadv2}
\end{minipage}
\end{table*}

We deduce from Table~\ref{tab-compcostsadv2}, that again, Euler forward is 
very inefficient, whereas for all higher-order schemes, the required grid
spacing is of similar orders of magnitude. Since the error is dominated by the
spatial one and the smoothness of the solution, increasing the Courant number
does not considerably decrease the accuracy. Once more, SSP\,RK(3,2) with
$\sigma=0.5$ turns out to be more efficient than TVD2 with $\sigma=0.25$.

In contrast to the advection equation where stability limits for $\sigma$ must 
be found by experiment, there are analytical methods to determine the maximum 
admissible Courant number for each time integration scheme. From Table~10 in 
\citet{KupkaHappenhoferHiguerasKoch2012} we deduce that the maximum Courant
number $\sigma$ as defined in equation~\eqref{eq-courdiff} for diffusive terms
is $0.187$ for Euler forward, $0.375$ for TVD2, $0.299$ for TVD3 and $0.672$
for SSP\,RK(3,2). For testing the efficiency of the diffusion equation, we use
the maximum allowed Courant numbers from
\citet{KupkaHappenhoferHiguerasKoch2012} for each scheme multiplied by $3/4$ due
to the stability constraint given in equation~(48) in
\citet{HappenhoferGrimm-StreleKupkaetal2013}. For the diffusion equation,
we expect much smaller errors due to the smoothing properties of the diffusion
equation. Therefore, we calculate the computational costs for a relative
accuracy of $1.0 \cdot 10^{-6}$.

\begin{table}[ht]
\centering
\begin{tabular}{c|rrrr}
scheme      & Euler & TVD2  & TVD3  & SSP\,RK(3,2) \\[1mm]
$\sigma$    & 0.187 & 0.375 & 0.299 & 0.672        \\
\hline                                                            \\[-3mm]  
$\varepsilon_{\rm rel}$ & \multicolumn{4}{c}{$1.0 \cdot 10^{-6}$} \\[1mm]
$h$                     & 4.19e-3 & 3.24e-2 & 3.65e-2 & 3.15e-2   \\[1mm] 
$N_h$                   &  1.36e7 &  2.95e4 &  2.06e4 &  3.19e4   \\[1mm] 
$N_t^{\star}$           &  4.06e5 &  6.79e3 &  1.00e4 &  5.98e3   \\[1mm] 
$\gamma^{\star}$        & 5.52e12 &  2.00e8 &  2.07e8 &  1.91e8 
\end{tabular}
\caption{Computational costs for diffusive problem with smooth initial condition.}
\label{tab-compcostsdiff}
\end{table}

The computational costs of the Euler forward scheme exceed the costs of the 
higher order schemes by several magnitudes. The costs with SSP\,RK(3,2) are 
significantly smaller than with TVD2 and TVD3, even though the advantage is 
smaller than one would expect from the difference in maximum allowed 
Courant numbers. 
 
In conclusion, for both advection and diffusion equations the WENO5 and the 
fourth order conservative finite difference scheme described in 
\citet{HappenhoferGrimm-StreleKupkaetal2013}, applied to the advection and diffusion
operator, respectively, combined with SSP\,RK(3,2) time integration are more efficient 
and more accurate than combinations with any other time integration schemes 
tested, both for smooth and non-smooth flows. We benefit from the high stability of 
SSP\,RK(3,2) and from the fact that {\em on grid sizes affordable in practice,}
the spatial error usually is much larger than the temporal one. This justifies 
the additional efforts required for the implementation of SSP\,RK(3,2), even though 
its theoretical order of accuracy is lower than TVD3 and it has more stages than 
TVD2.

Changes of the time-integration scheme do not affect the number of grid
cell updates per CPU second performed by the code, but they do change the
overall number of updates needed for the simulation of a fixed time span. From
Tables~\ref{tab-compcostsadv1} and~\ref{tab-compcostsdiff} we conclude that the
achievable change of required grid cell updates is of the order of several tens
of percent. This number directly translates into a speedup in terms of
computational time, if the spatial discretisation is not changed.

We emphasize that these results do not tell that lower order schemes are 
{\em in general} more efficient than higher order schemes. In fact, our analysis
only applied to simulations of solar surface convection with the specific 
numerical schemes we used. In particular, there might be other Runge--Kutta
schemes (as, e.g., the ones presented in \citet{Ketcheson2008}) which 
are even more efficient. Nevertheless, our analysis provides a valuable tool
to assess the achievable efficiency with a certain combination of numerical 
schemes and for a specific application.

\section*{Acknowledgements}
We acknowledge financial support from the Austrian Science fund (FWF), projects 
P25229 and P21742. HGS wants to thank the MPA Garching for 
a grant for a research stay in Garching. Calculations have been performed at the
VSC clusters of the Vienna universities.

\appendix



\section{Error Sizes for the Advection Equation with Smooth Initial Data}
\label{app-advsin}

\begin{table*}[ht]
\centering
\begin{tabular}{l|llllllllllll}
$\delta x$ \textbackslash\ $\delta t$ & 0.1250 & 0.0625 & 0.0312 & 0.0156 & 0.0078 & 0.0039 & 0.0020 & 0.0010 & 0.0005 & 0.0002 & 0.0001 \\ 
\hline
0.1250 &         & 4.26e-1 & 1.07e-1 & 3.39e-2 & 9.93e-3 & 7.07e-3 & 1.00e-2 & 1.18e-2 & 1.27e-2 & 1.32e-2 & 1.34e-2 \\ 
0.0625 &         &         & 1.33e-1 & 5.66e-2 & 2.37e-2 & 1.06e-2 & 4.74e-3 & 2.01e-3 & 7.66e-4 & 4.62e-4 & 5.96e-4 \\ 
0.0312 &         &         &         & 4.14e-2 & 2.48e-2 & 1.15e-2 & 5.55e-3 & 2.71e-3 & 1.33e-3 & 6.46e-4 & 3.09e-4 \\ 
0.0156 &         &         &         &         & 1.57e-2 & 1.00e-2 & 5.62e-3 & 2.76e-3 & 1.36e-3 & 6.79e-4 & 3.38e-4 \\ 
0.0078 &         &         &         &         &         & 7.87e-3 & 4.24e-3 & 2.77e-3 & 1.37e-3 & 6.82e-4 & 3.40e-4 \\ 
0.0039 &         &         &         &         &         &         & 4.01e-3 & 1.95e-3 & 1.39e-3 & 6.84e-4 & 3.41e-4 \\ 
0.0020 &         &         &         &         &         &         &         & 3.87e-3 & 1.60e-3 & 5.94e-4 & 3.41e-4 \\ 
0.0010 &         &         &         &         &         &         &         &         & 1.36e-2 & 1.41e-3 & 3.87e-4 
\end{tabular}
\caption{$\varepsilon(\delta x,\delta t)$ for the combination of WENO5 with Euler forward
         when solving~\eqref{eq-adv1d} \& \eqref{eq-sinic}.}
\label{tab-adv1Euler}

\vspace*{1cm}

\begin{tabular}{l|llllllllllll}
$\delta x$ \textbackslash\ $\delta t$ & 0.1250 & 0.0625 & 0.0312 & 0.0156 & 0.0078 & 0.0039 & 0.0020 & 0.0010 & 0.0005 & 0.0002 & 0.0001 \\ 
\hline
0.1250 & 7.68e-2 & 1.53e-2 & 1.21e-2 & 1.33e-2 & 1.36e-2 & 1.36e-2 & 1.37e-2 & 1.37e-2 & 1.37e-2 & 1.37e-2 & 1.37e-2 \\ 
0.0625 &         & 2.21e-2 & 5.39e-3 & 1.47e-3 & 8.55e-4 & 8.35e-4 & 8.40e-4 & 8.43e-4 & 8.43e-4 & 8.43e-4 & 8.43e-4 \\ 
0.0312 &         &         & 6.10e-3 & 1.41e-3 & 3.50e-4 & 9.03e-5 & 3.64e-5 & 3.10e-5 & 3.09e-5 & 3.10e-5 & 3.10e-5 \\ 
0.0156 &         &         &         & 6.78e-3 & 3.55e-4 & 8.86e-5 & 2.21e-5 & 5.58e-6 & 1.65e-6 & 1.01e-6 & 9.65e-7 \\ 
0.0078 &         &         &         &         & 4.28e-3 & 8.90e-5 & 2.22e-5 & 5.56e-6 & 1.39e-6 & 3.48e-7 & 9.12e-8 \\ 
0.0039 &         &         &         &         &         & 2.26e-3 & 2.23e-5 & 5.57e-6 & 1.39e-6 & 3.48e-7 & 8.69e-8 \\ 
0.0020 &         &         &         &         &         &         & 1.09e-3 & 5.57e-6 & 1.39e-6 & 3.48e-7 & 8.70e-8 \\ 
0.0010 &         &         &         &         &         &         &         & 5.86e-4 & 1.39e-6 & 3.48e-7 & 8.71e-8 
\end{tabular}
\caption{$\varepsilon(\delta x,\delta t)$ for the combination of WENO5 with TVD2
         when solving~\eqref{eq-adv1d} \& \eqref{eq-sinic}.}
\label{tab-adv1TVD2}
\end{table*}

\begin{table*}[ht]
\centering
\begin{tabular}{l|llllllllllll}
$\delta x$ \textbackslash\ $\delta t$ & 0.1250 & 0.0625 & 0.0312 & 0.0156 & 0.0078 & 0.0039 & 0.0020 & 0.0010 & 0.0005 & 0.0002 & 0.0001 \\ 
\hline
0.1250 & 2.19e-2 & 1.54e-2 & 1.38e-2 & 1.38e-2 & 1.37e-2 & 1.37e-2 & 1.37e-2 & 1.37e-2 & 1.37e-2 &  1.37e-2 &  1.37e-2 \\ 
0.0625 &         & 2.89e-3 & 1.06e-3 & 8.75e-4 & 8.49e-4 & 8.45e-4 & 8.43e-4 & 8.44e-4 & 8.43e-4 &  8.43e-4 &  8.43e-4 \\ 
0.0312 &         &         & 3.07e-4 & 6.38e-5 & 3.49e-5 & 3.14e-5 & 3.11e-5 & 3.10e-5 & 3.10e-5 &  3.10e-5 &  3.10e-5 \\ 
0.0156 &         &         &         & 3.59e-5 & 5.29e-6 & 1.49e-6 & 1.03e-6 & 9.73e-7 & 9.66e-7 &  9.65e-7 &  9.65e-7 \\ 
0.0078 &         &         &         &         & 4.41e-6 & 5.76e-7 & 9.77e-8 & 3.83e-8 & 3.09e-8 &  3.00e-8 &  2.99e-8 \\ 
0.0039 &         &         &         &         &         & 5.48e-7 & 6.93e-8 & 9.46e-9 & 1.99e-9 &  1.06e-9 & 9.44e-10 \\ 
0.0020 &         &         &         &         &         &         & 6.85e-8 & 8.58e-9 & 1.10e-9 & 1.62e-10 & 4.56e-11 \\ 
0.0010 &         &         &         &         &         &         &         & 8.55e-9 & 1.07e-9 & 1.34e-10 & 1.76e-11 
\end{tabular}
\caption{$\varepsilon(\delta x,\delta t)$ for the combination of WENO5 with TVD3
         when solving~\eqref{eq-adv1d} \& \eqref{eq-sinic}.}
\label{tab-adv1TVD3}

\vspace*{1cm}

\begin{tabular}{l|llllllllllll}
$\delta x$ \textbackslash\ $\delta t$ & 0.1250 & 0.0625 & 0.0312 & 0.0156 & 0.0078 & 0.0039 & 0.0020 & 0.0010 & 0.0005 & 0.0002 & 0.0001 \\ 
\hline
0.1250 & 3.55e-2 & 1.23e-2 & 1.28e-2 & 1.36e-2 & 1.37e-2 & 1.37e-2 & 1.37e-2 & 1.37e-2 & 1.37e-2 & 1.37e-2 & 1.37e-2 \\ 
0.0625 & 1.58e-1 & 1.14e-2 & 2.71e-3 & 9.91e-4 & 8.34e-4 & 8.39e-4 & 8.42e-4 & 8.43e-4 & 8.43e-4 & 8.43e-4 & 8.43e-4 \\ 
0.0312 &         & 9.20e-2 & 2.86e-3 & 7.05e-4 & 1.76e-4 & 5.14e-5 & 3.20e-5 & 3.09e-5 & 3.09e-5 & 3.10e-5 & 3.10e-5 \\ 
0.0156 &         &         &         & 7.14e-4 & 1.78e-4 & 4.43e-5 & 1.11e-5 & 2.89e-6 & 1.16e-6 & 9.73e-7 & 9.64e-7 \\ 
0.0078 &         &         &         & 8.09e-2 & 1.79e-4 & 4.45e-5 & 1.11e-5 & 2.78e-6 & 6.94e-7 & 1.76e-7 & 5.22e-8 \\ 
0.0039 &         &         &         &         &         & 4.06e-4 & 1.11e-5 & 2.78e-6 & 6.96e-7 & 1.74e-7 & 4.35e-8 \\ 
0.0020 &         &         &         &         &         &         & 2.49e-4 & 2.79e-6 & 6.96e-7 & 1.74e-7 & 4.35e-8 \\ 
0.0010 &         &         &         &         &         &         &         & 1.55e-4 & 6.97e-7 & 1.74e-7 & 4.35e-8 
\end{tabular}
\caption{$\varepsilon(\delta x,\delta t)$ for the combination of WENO5 with SSP\,RK(3,2)
         when solving~\eqref{eq-adv1d} \& \eqref{eq-sinic}.}
\label{tab-adv1SSPRK32}
\end{table*}

\section{Error Sizes for the Advection Equation with Discontinuous Initial Data}
\label{app-advstp}

\begin{table*}[ht]
\centering
\begin{tabular}{l|llllllllllll}
$\delta x$ \textbackslash\ $\delta t$ & 0.1250 & 0.0625 & 0.0312 & 0.0156 & 0.0078 & 0.0039 & 0.0020 & 0.0010 & 0.0005 & 0.0002 & 0.0001 \\ 
\hline
0.1250 &         & 2.52e-1 & 3.75e-1 & 3.38e-1 & 3.45e-1 & 3.53e-1 & 3.53e-1 & 3.55e-1 & 3.56e-1 & 3.56e-1 & 3.56e-1 \\ 
0.0625 &         & 2.05e-3 & 2.67e-1 & 3.46e-1 & 2.32e-1 & 2.27e-1 & 2.38e-1 & 2.42e-1 & 2.46e-1 & 2.47e-1 & 2.48e-1 \\ 
0.0312 &         &         &         & 1.24e-1 & 3.41e-1 & 2.25e-1 & 1.54e-1 & 1.40e-1 & 1.44e-1 & 1.48e-1 & 1.50e-1 \\ 
0.0156 &         &         &         &         & 8.94e-2 & 1.02e-1 & 1.13e-1 & 9.45e-2 & 1.05e-1 & 1.12e-1 & 1.17e-1 \\ 
0.0078 &         &         &         &         &         & 6.43e-2 & 5.04e-2 & 5.79e-2 & 6.61e-2 & 7.39e-2 & 8.05e-2 \\ 
0.0039 &         &         &         &         &         &         & 5.28e-2 & 3.63e-2 & 4.09e-2 & 4.68e-2 & 5.23e-2 \\ 
0.0020 &         &         &         &         &         &         &         & 1.61e-1 & 1.85e-1 & 2.89e-2 & 3.30e-2 \\ 
0.0010 &         &         &         &         &         &         &         &         & 9.86e-1 &         & 2.26e-2 
\end{tabular}
\caption{$\varepsilon(\delta x,\delta t)$ for the combination of WENO5 with Euler forward
         when solving~\eqref{eq-adv1d} \& \eqref{eq-stepic}.}
\label{tab-adv2Euler}

\vspace*{1cm}

\begin{tabular}{l|llllllllllll}
$\delta x$ \textbackslash\ $\delta t$ & 0.1250 & 0.0625 & 0.0312 & 0.0156 & 0.0078 & 0.0039 & 0.0020 & 0.0010 & 0.0005 & 0.0002 & 0.0001 \\ 
\hline
0.1250 & 4.64e-1 & 3.22e-1 & 3.49e-1 & 3.47e-1 & 3.52e-1 & 3.56e-1 & 3.55e-1 & 3.56e-1 & 3.56e-1 & 3.56e-1 & 3.56e-1 \\ 
0.0625 &         & 3.67e-1 & 2.52e-1 & 2.38e-1 & 2.43e-1 & 2.46e-1 & 2.49e-1 & 2.48e-1 & 2.49e-1 & 2.49e-1 & 2.49e-1 \\ 
0.0312 &         &         & 3.13e-1 & 1.99e-1 & 1.58e-1 & 1.52e-1 & 1.52e-1 & 1.52e-1 & 1.52e-1 & 1.52e-1 & 1.52e-1 \\ 
0.0156 &         &         &         & 2.48e-1 & 1.38e-1 & 1.29e-1 & 1.32e-1 & 1.21e-1 & 1.21e-1 & 1.21e-1 & 1.21e-1 \\ 
0.0078 &         &         &         &         & 1.93e-1 & 1.06e-1 & 9.20e-2 & 9.02e-2 & 9.00e-2 & 9.00e-2 & 9.00e-2 \\ 
0.0039 &         &         &         &         &         & 1.58e-1 & 7.92e-2 & 6.98e-2 & 6.77e-2 & 6.76e-2 & 6.76e-2 \\ 
0.0020 &         &         &         &         &         &         & 1.30e-1 & 6.75e-2 & 5.36e-2 & 5.08e-2 & 5.07e-2 \\ 
0.0010 &         &         &         &         &         &         &         & 1.09e-1 & 5.27e-2 & 4.17e-2 & 3.80e-2 
\end{tabular}
\caption{$\varepsilon(\delta x,\delta t)$ for the combination of WENO5 with TVD2
         when solving~\eqref{eq-adv1d} \& \eqref{eq-stepic}.}
\label{tab-adv2TVD2}
\end{table*}

\begin{table*}[ht]
\centering
\begin{tabular}{l|llllllllllll}
$\delta x$ \textbackslash\ $\delta t$ & 0.1250 & 0.0625 & 0.0312 & 0.0156 & 0.0078 & 0.0039 & 0.0020 & 0.0010 & 0.0005 & 0.0002 & 0.0001 \\ 
\hline
0.1250 & 3.66e-1 & 3.38e-1 & 3.56e-1 & 3.49e-1 & 3.53e-1 & 3.56e-1 & 3.55e-1 & 3.56e-1 & 3.56e-1 & 3.56e-1 & 3.56e-1 \\ 
0.0625 &         & 3.28e-1 & 2.47e-1 & 2.42e-1 & 2.45e-1 & 2.47e-1 & 2.49e-1 & 2.48e-1 & 2.49e-1 & 2.49e-1 & 2.49e-1 \\ 
0.0312 &         &         & 2.47e-1 & 1.65e-1 & 1.55e-1 & 1.52e-1 & 1.52e-1 & 1.52e-1 & 1.52e-1 & 1.52e-1 & 1.52e-1 \\ 
0.0156 &         &         &         & 1.92e-1 & 1.30e-1 & 1.30e-1 & 1.32e-1 & 1.21e-1 & 1.21e-1 & 1.21e-1 & 1.21e-1 \\ 
0.0078 &         &         &         &         & 1.74e-1 & 9.46e-2 & 9.11e-2 & 9.03e-2 & 9.00e-2 & 9.00e-2 & 9.00e-2 \\ 
0.0039 &         &         &         &         &         & 1.48e-1 & 7.06e-2 & 6.83e-2 & 6.78e-2 & 6.76e-2 & 6.76e-2 \\ 
0.0020 &         &         &         &         &         &         & 1.20e-1 & 5.29e-2 & 5.11e-2 & 5.08e-2 & 5.07e-2 \\ 
0.0010 &         &         &         &         &         &         &         & 1.01e-1 & 3.96e-2 & 3.82e-2 & 3.80e-2 
\end{tabular}
\caption{$\varepsilon(\delta x,\delta t)$ for the combination of WENO5 with TVD3
         when solving~\eqref{eq-adv1d} \& \eqref{eq-stepic}.}
\label{tab-adv2TVD3}

\vspace*{1cm}

\begin{tabular}{l|llllllllllll}
$\delta x$ \textbackslash\ $\delta t$ & 0.1250 & 0.0625 & 0.0312 & 0.0156 & 0.0078 & 0.0039 & 0.0020 & 0.0010 & 0.0005 & 0.0002 & 0.0001 \\ 
\hline
0.1250 & 3.49e-1 & 3.31e-1 & 3.53e-1 & 3.48e-1 & 3.52e-1 & 3.56e-1 & 3.55e-1 & 3.56e-1 & 3.56e-1 & 3.56e-1 & 3.56e-1 \\ 
0.0625 & 4.48e-1 & 3.01e-1 & 2.36e-1 & 2.40e-1 & 2.44e-1 & 2.46e-1 & 2.49e-1 & 2.48e-1 & 2.49e-1 & 2.49e-1 & 2.49e-1 \\ 
0.0312 &         &         & 1.99e-1 & 1.77e-1 & 1.56e-1 & 1.52e-1 & 1.52e-1 & 1.52e-1 & 1.52e-1 & 1.52e-1 & 1.52e-1 \\ 
0.0156 &         &         &         & 1.54e-1 & 1.29e-1 & 1.29e-1 & 1.32e-1 & 1.21e-1 & 1.21e-1 & 1.21e-1 & 1.21e-1 \\ 
0.0078 &         &         &         &         & 1.18e-1 & 9.58e-2 & 9.13e-2 & 9.02e-2 & 9.00e-2 & 9.00e-2 & 9.00e-2 \\ 
0.0039 &         &         &         &         &         & 9.33e-2 & 6.99e-2 & 6.86e-2 & 6.77e-2 & 6.76e-2 & 6.76e-2 \\ 
0.0020 &         &         &         &         &         &         & 7.42e-2 & 6.08e-2 & 5.17e-2 & 5.08e-2 & 5.07e-2 \\ 
0.0010 &         &         &         &         &         &         &         & 6.28e-2 & 4.75e-2 & 3.93e-2 & 3.79e-2 
\end{tabular}
\caption{$\varepsilon(\delta x,\delta t)$ for the combination of WENO5 with SSP\,RK(3,2)
         when solving~\eqref{eq-adv1d} \& \eqref{eq-stepic}.}
\label{tab-adv2SSPRK32}
\end{table*}

\section{Error Sizes for the Diffusion Equation with Smooth Initial Data}
\label{app-diff}

\begin{table*}[ht]
\centering
\begin{tabular}{l|lllllllllll}
$\delta x$ \textbackslash\ $\delta t$ & 1.95312 & 0.48828 & 0.24414 & 0.12207 & 0.06104 & 0.03052 & 0.01526 & 0.00763 & 0.00191 & 0.00048 & 0.00012 \\ 
\hline
0.12500 & 6.92e-4 & 1.25e-4 & 3.14e-5 & 1.54e-5 & 3.88e-5 & 5.06e-5 & 5.64e-5 & 5.93e-5 & 6.15e-5 & 6.21e-5 & 6.22e-5 \\ 
0.06250 & 7.56e-4 & 1.85e-4 & 9.03e-5 & 4.28e-5 & 1.91e-5 & 7.29e-6 & 1.37e-6 & 1.58e-6 & 3.80e-6 & 4.36e-6 & 4.50e-6 \\ 
0.03125 &         &         & 9.42e-5 & 4.69e-5 & 2.33e-5 & 1.15e-5 & 5.60e-6 & 2.65e-6 & 4.34e-7 & 1.19e-7 & 2.57e-7 \\ 
0.01562 &         &         &         &         & 2.35e-5 & 1.17e-5 & 5.86e-6 & 2.92e-6 & 7.15e-7 & 1.64e-7 & 2.64e-8 \\ 
0.00781 &         &         &         &         &         &         & 5.86e-6 & 2.93e-6 & 7.31e-7 & 1.82e-7 & 4.45e-8 \\ 
0.00391 &         &         &         &         &         &         &         &         & 7.31e-7 & 1.83e-7 & 4.56e-8 \\ 
0.00195 &         &         &         &         &         &         &         &         &         & 1.83e-7 & 4.57e-8 \\ 
0.00098 &         &         &         &         &         &         &         &         &         &         & 4.56e-8 
\end{tabular}
\caption{$\varepsilon(\delta x,\delta t)$ for the combination of WENO5 with Euler forward
         when solving~\eqref{eq-diff1d} \& \eqref{diff-sinic}.}
\label{tab-diffEuler}

\vspace*{1cm}

\begin{tabular}{l|lllllllllll}
$\delta x$ \textbackslash\ $\delta t$ & 1.95312 & 0.48828 & 0.24414 & 0.12207 & 0.06104 & 0.03052 & 0.01526 & 0.00763 & 0.00191 & 0.00048 & 0.00012 \\ 
\hline
0.12500 & 7.74e-5 & 6.30e-5 & 6.25e-5 & 6.23e-5 & 6.22e-5 & 6.23e-5 & 6.23e-5 & 6.23e-5 &  6.23e-5 &  6.23e-5 &  6.23e-5 \\ 
0.06250 & 2.24e-5 & 5.61e-6 & 4.81e-6 & 4.61e-6 & 4.56e-6 & 4.55e-6 & 4.54e-6 & 4.54e-6 &  4.54e-6 &  4.54e-6 &  4.54e-6 \\ 
0.03125 &         &         & 5.90e-7 & 3.74e-7 & 3.21e-7 & 3.08e-7 & 3.04e-7 & 3.03e-7 &  3.03e-7 &  3.03e-7 &  3.03e-7 \\ 
0.01562 &         &         &         &         & 3.78e-8 & 2.41e-8 & 2.06e-8 & 1.98e-8 &  1.95e-8 &  1.95e-8 &  1.95e-8 \\ 
0.00781 &         &         &         &         &         &         & 2.39e-9 & 1.52e-9 &  1.25e-9 &  1.24e-9 &  1.24e-9 \\ 
0.00391 &         &         &         &         &         &         &         &         & 9.59e-11 & 8.01e-11 & 7.78e-11 \\ 
0.00195 &         &         &         &         &         &         &         &         &          & 6.80e-12 & 4.81e-12 \\ 
0.00098 &         &         &         &         &         &         &         &         &          &          & 5.06e-13  
\end{tabular}
\caption{$\varepsilon(\delta x,\delta t)$ for the combination of WENO5 with TVD2
         when solving~\eqref{eq-diff1d} \& \eqref{diff-sinic}.}
\label{tab-diffTVD2}
\end{table*}

\begin{table*}[ht]
\centering
\begin{tabular}{l|lllllllllll}
$\delta x$ \textbackslash\ $\delta t$ & 1.95312 & 0.48828 & 0.24414 & 0.12207 & 0.06104 & 0.03052 & 0.01526 & 0.00763 & 0.00191 & 0.00048 & 0.00012 \\ 
\hline
0.12500 & 6.15e-5 & 6.21e-5 & 6.22e-5 & 6.22e-5 & 6.22e-5 & 6.22e-5 & 6.23e-5 & 6.23e-5 &  6.23e-5 &  6.23e-5 &  6.23e-5 \\ 
0.06250 & 4.18e-6 & 4.52e-6 & 4.54e-6 & 4.54e-6 & 4.54e-6 & 4.54e-6 & 4.54e-6 & 4.54e-6 &  4.54e-6 &  4.54e-6 &  4.54e-6 \\ 
0.03125 &         & 9.61e-7 & 3.02e-7 & 3.03e-7 & 3.03e-7 & 3.03e-7 & 3.03e-7 & 3.03e-7 &  3.03e-7 &  3.03e-7 &  3.03e-7 \\ 
0.01562 &         &         &         &         & 1.95e-8 & 1.95e-8 & 1.95e-8 & 1.95e-8 &  1.95e-8 &  1.95e-8 &  1.95e-8 \\ 
0.00781 &         &         &         &         &         &         & 1.23e-9 & 1.24e-9 &  1.24e-9 &  1.24e-9 &  1.24e-9 \\ 
0.00391 &         &         &         &         &         &         &         &         & 7.77e-11 & 7.79e-11 & 8.17e-11 \\ 
0.00195 &         &         &         &         &         &         &         &         &          & 8.04e-12 & 2.58e-11 \\ 
0.00098 &         &         &         &         &         &         &         &         &          &          & 2.56e-11  
\end{tabular}
\caption{$\varepsilon(\delta x,\delta t)$ for the combination of WENO5 with TVD3
         when solving~\eqref{eq-diff1d} \& \eqref{diff-sinic}.}
\label{tab-diffTVD3}

\vspace*{1cm}

\begin{tabular}{l|lllllllllll}
$\delta x$ \textbackslash\ $\delta t$ & 1.95312 & 0.48828 & 0.24414 & 0.12207 & 0.06104 & 0.03052 & 0.01526 & 0.00763 & 0.00191 & 0.00048 & 0.00012 \\ 
\hline
0.12500 & 6.95e-5 & 6.25e-5 & 6.23e-5 & 6.23e-5 & 6.22e-5 & 6.23e-5 & 6.23e-5 &  6.23e-5 &  6.23e-5 &  6.23e-5 &  6.23e-5 \\ 
0.06250 & 1.33e-5 & 5.06e-6 & 4.68e-6 & 4.57e-6 & 4.55e-6 & 4.54e-6 & 4.54e-6 &  4.54e-6 &  4.54e-6 &  4.54e-6 &  4.54e-6 \\ 
0.03125 &         & 8.74e-7 & 4.46e-7 & 3.39e-7 & 3.12e-7 & 3.05e-7 & 3.04e-7 &  3.03e-7 &  3.03e-7 &  3.03e-7 &  3.03e-7 \\ 
0.01562 &         &         &         & 5.61e-8 & 2.86e-8 & 2.18e-8 & 2.01e-8 &  1.96e-8 &  1.95e-8 &  1.95e-8 &  1.95e-8 \\ 
0.00781 &         &         &         &         &         & 3.55e-9 & 1.81e-9 &  1.38e-9 &  1.24e-9 &  1.24e-9 &  1.24e-9 \\ 
0.00391 &         &         &         &         &         &         &         & 2.23e-10 & 8.68e-11 & 7.86e-11 & 8.26e-11 \\ 
0.00195 &         &         &         &         &         &         &         &          & 1.41e-11 & 8.65e-12 & 2.84e-11 \\ 
0.00098 &         &         &         &         &         &         &         &          &          & 6.55e-12 & 2.73e-11  
\end{tabular}
\caption{$\varepsilon(\delta x,\delta t)$ for the combination of WENO5 with SSP\,RK(3,2)
         when solving~\eqref{eq-diff1d} \& \eqref{diff-sinic}.}
\label{tab-diffSSPRK32}
\end{table*}




\bibliographystyle{model1-num-names}

\begin{thebibliography}{33}
\expandafter\ifx\csname natexlab\endcsname\relax\def\natexlab#1{#1}\fi
\providecommand{\url}[1]{\texttt{#1}}
\providecommand{\href}[2]{#2}
\providecommand{\path}[1]{#1}
\providecommand{\DOIprefix}{doi:}
\providecommand{\ArXivprefix}{arXiv:}
\providecommand{\URLprefix}{URL: }
\providecommand{\Pubmedprefix}{pmid:}
\providecommand{\doi}[1]{\href{http://dx.doi.org/#1}{\path{#1}}}
\providecommand{\Pubmed}[1]{\href{pmid:#1}{\path{#1}}}
\providecommand{\bibinfo}[2]{#2}
\ifx\xfnm\relax \def\xfnm[#1]{\unskip,\space#1}\fi
\bibitem[{{M}uthsam et~al.(2010){M}uthsam, {K}upka, {L\"ow-Baselli},
  {O}bertscheider, {L}anger, and {L}enz}]{MuthsamKupkaLoew-Basellietal2010}
\bibinfo{author}{H.~J. {M}uthsam}, \bibinfo{author}{F.~{K}upka},
  \bibinfo{author}{B.~{L\"ow-Baselli}}, \bibinfo{author}{C.~{O}bertscheider},
  \bibinfo{author}{M.~{L}anger}, \bibinfo{author}{P.~{L}enz},
\newblock \bibinfo{journal}{NewA} \bibinfo{volume}{15} (\bibinfo{year}{2010})
  \bibinfo{pages}{460 -- 475}.
\bibitem[{{M}undprecht et~al.(2013){M}undprecht, {M}uthsam, and
  {K}upka}]{MundprechtMuthsamKupka2013}
\bibinfo{author}{E.~{M}undprecht}, \bibinfo{author}{H.~J. {M}uthsam},
  \bibinfo{author}{F.~{K}upka},
\newblock \bibinfo{journal}{MNRAS} \bibinfo{volume}{435} (\bibinfo{year}{2013})
  \bibinfo{pages}{3191 -- 3205}.
\bibitem[{{Z}aussinger and {S}pruit(2013)}]{ZaussingerSpruit2013}
\bibinfo{author}{F.~{Z}aussinger}, \bibinfo{author}{H.~{S}pruit},
\newblock \bibinfo{journal}{A\&A} \bibinfo{volume}{554} (\bibinfo{year}{2013})
  \bibinfo{pages}{A119}.
\bibitem[{{T}oro(2009)}]{Toro2009}
\bibinfo{author}{E.~F. {T}oro}, \bibinfo{title}{{R}iemann solvers and numerical
  methods for fluid dynamics: a practical introduction},
  \bibinfo{publisher}{Springer Berlin New--York Heidelberg},
  \bibinfo{year}{2009}.
\bibitem[{{L}e{V}eque(2007)}]{LeVeque2007}
\bibinfo{author}{R.~J. {L}e{V}eque}, \bibinfo{title}{{F}inite {D}ifference
  {M}ethods for {O}rdinary and {P}artial {D}ifferential {E}quations. {S}teady
  {S}tate and {T}ime {D}ependent {P}roblems}, \bibinfo{publisher}{Society for
  Industrial and Applied Mathematics (SIAM)}, \bibinfo{year}{2007}.
\bibitem[{{K}upka et~al.(2012){K}upka, {H}appenhofer, {H}igueras, and
  {K}och}]{KupkaHappenhoferHiguerasKoch2012}
\bibinfo{author}{F.~{K}upka}, \bibinfo{author}{N.~{H}appenhofer},
  \bibinfo{author}{I.~{H}igueras}, \bibinfo{author}{O.~{K}och},
\newblock \bibinfo{journal}{JCP} \bibinfo{volume}{231} (\bibinfo{year}{2012})
  \bibinfo{pages}{3561 -- 3586}.
\bibitem[{{S}hu and {O}sher(1988)}]{ShuOsher1988}
\bibinfo{author}{C.-W. {S}hu}, \bibinfo{author}{S.~{O}sher},
\newblock \bibinfo{journal}{JCP} \bibinfo{volume}{77} (\bibinfo{year}{1988})
  \bibinfo{pages}{439 -- 471}.
\bibitem[{{S}hu(2003)}]{Shu2003}
\bibinfo{author}{C.-W. {S}hu},
\newblock \bibinfo{journal}{International Journal of Computational Fluid
  Dynamics} \bibinfo{volume}{17} (\bibinfo{year}{2003}) \bibinfo{pages}{107 --
  118}.
\bibitem[{{M}erriman(2003)}]{Merriman2003}
\bibinfo{author}{B.~{M}erriman},
\newblock \bibinfo{journal}{Journal Of Scientific Computing}
  \bibinfo{volume}{19} (\bibinfo{year}{2003}) \bibinfo{pages}{309 -- 322}.
\bibitem[{{M}uthsam et~al.(2007){M}uthsam, {L\"ow-Baselli}, {O}bertscheider,
  {L}anger, {L}enz, and {K}upka}]{MuthsamLoew-BaselliOberscheideretal2007}
\bibinfo{author}{H.~J. {M}uthsam}, \bibinfo{author}{B.~{L\"ow-Baselli}},
  \bibinfo{author}{C.~{O}bertscheider}, \bibinfo{author}{M.~{L}anger},
  \bibinfo{author}{P.~{L}enz}, \bibinfo{author}{F.~{K}upka},
\newblock \bibinfo{journal}{MNRAS} \bibinfo{volume}{380} (\bibinfo{year}{2007})
  \bibinfo{pages}{1335 -- 1340}.
\bibitem[{{H}appenhofer et~al.(2013){H}appenhofer, {Grimm-Strele}, {K}upka,
  {L\"ow-Baselli}, and {M}uthsam}]{HappenhoferGrimm-StreleKupkaetal2013}
\bibinfo{author}{N.~{H}appenhofer}, \bibinfo{author}{H.~{Grimm-Strele}},
  \bibinfo{author}{F.~{K}upka}, \bibinfo{author}{B.~{L\"ow-Baselli}},
  \bibinfo{author}{H.~J. {M}uthsam},
\newblock \bibinfo{journal}{JCP} \bibinfo{volume}{236} (\bibinfo{year}{2013})
  \bibinfo{pages}{96 -- 118}.
\bibitem[{{G}ottlieb et~al.(2001){G}ottlieb, {S}hu, and
  {T}admor}]{GottliebShuTadmor2001}
\bibinfo{author}{S.~{G}ottlieb}, \bibinfo{author}{C.-W. {S}hu},
  \bibinfo{author}{E.~{T}admor},
\newblock \bibinfo{journal}{SIAM Review} \bibinfo{volume}{43}
  (\bibinfo{year}{2001}) \bibinfo{pages}{89 -- 112}.
\bibitem[{{W}ang and {S}piteri(2007)}]{WangSpiteri2007}
\bibinfo{author}{R.~{W}ang}, \bibinfo{author}{R.~J. {S}piteri},
\newblock \bibinfo{journal}{SIAM J. Numer. Anal.} \bibinfo{volume}{45}
  (\bibinfo{year}{2007}) \bibinfo{pages}{1871 -- 1901}.
\bibitem[{{K}raaijevanger(1991)}]{Kraaijevanger1991}
\bibinfo{author}{J.~F.~B.~M. {K}raaijevanger},
\newblock \bibinfo{journal}{BIT} \bibinfo{volume}{31} (\bibinfo{year}{1991})
  \bibinfo{pages}{482 -- 528}.
\bibitem[{{K}etcheson et~al.(2009){K}etcheson, {M}acdonald, and
  {G}ottlieb}]{KetchesonMacdonaldGottlieb2009}
\bibinfo{author}{D.~I. {K}etcheson}, \bibinfo{author}{C.~B. {M}acdonald},
  \bibinfo{author}{S.~{G}ottlieb},
\newblock \bibinfo{journal}{Applied Numerical Mathematics} \bibinfo{volume}{59}
  (\bibinfo{year}{2009}) \bibinfo{pages}{373 -- 392}.
\bibitem[{{H}eun(1900)}]{Heun1900}
\bibinfo{author}{K.~{H}eun},
\newblock \bibinfo{journal}{Z. Math. Phys} \bibinfo{volume}{45}
  (\bibinfo{year}{1900}) \bibinfo{pages}{23 -- 38}.
\bibitem[{{F}ehlberg(1970)}]{Fehlberg1970}
\bibinfo{author}{E.~{F}ehlberg},
\newblock \bibinfo{journal}{Computing} \bibinfo{volume}{6}
  (\bibinfo{year}{1970}) \bibinfo{pages}{61 -- 71}.
\bibitem[{{K}etcheson(2008)}]{Ketcheson2008}
\bibinfo{author}{D.~I. {K}etcheson},
\newblock \bibinfo{journal}{SIAM J. Sci. Comput.} \bibinfo{volume}{30}
  (\bibinfo{year}{2008}) \bibinfo{pages}{2113--2136}.
\bibitem[{Happenhofer et~al.(2011)Happenhofer, Koch, and
  Kupka}]{HappenhoferKochKupka2011}
\bibinfo{author}{N.~Happenhofer}, \bibinfo{author}{O.~Koch},
  \bibinfo{author}{F.~Kupka}, 
  \bibinfo{type}{ASC Report} \bibinfo{number}{27}, Institute for
  Analysis and Scientific Computing, Vienna UT, \bibinfo{year}{2011}.
\bibitem[{{M}otamed et~al.(2011){M}otamed, {M}acdonald, and
  {R}uuth}]{MotamedMacdonaldRuuth2011}
\bibinfo{author}{M.~{M}otamed}, \bibinfo{author}{C.~B. {M}acdonald},
  \bibinfo{author}{S.~J. {R}uuth},
\newblock \bibinfo{journal}{Journal Of Scientific Computing}
  \bibinfo{volume}{47} (\bibinfo{year}{2011}) \bibinfo{pages}{127 -- 149}.
\bibitem[{{F}erziger and {P}eri\'{c}(2002)}]{FerzigerPeric2002}
\bibinfo{author}{J.~H. {F}erziger}, \bibinfo{author}{M.~{P}eri\'{c}},
  \bibinfo{title}{{C}omputational {M}ethods for {F}luid {D}ynamics},
  \bibinfo{edition}{3rd} ed., \bibinfo{publisher}{Springer},
  \bibinfo{address}{Berlin}, \bibinfo{year}{2002}.
\bibitem[{{E}vans(2002)}]{Evans2002}
\bibinfo{author}{L.~C. {E}vans}, \bibinfo{title}{{P}artial {D}ifferential
  {E}quations}, volume~\bibinfo{volume}{19} of
  \textit{\bibinfo{series}{Graduate Studies in Mathematics}},
  \bibinfo{edition}{2nd} ed., \bibinfo{publisher}{American Mathematical
  Society}, \bibinfo{year}{2002}.
\bibitem[{{S}trikwerda(1989)}]{Strikwerda1989}
\bibinfo{author}{J.~C. {S}trikwerda}, \bibinfo{title}{{F}inite {D}ifference
  {S}chemes and {P}artial {D}ifferential {E}quations},
  \bibinfo{publisher}{Wadsworth \& Brooks/Cole}, \bibinfo{year}{1989}.
\bibitem[{{S}magorinsky(1963)}]{Smagorinsky1963}
\bibinfo{author}{J.~{S}magorinsky},
\newblock \bibinfo{journal}{MWR} \bibinfo{volume}{91} (\bibinfo{year}{1963})
  \bibinfo{pages}{99 -- 164}.
\bibitem[{{Grimm-Strele} et~al.(2015){Grimm-Strele}, {K}upka, {L\"ow-Baselli},
  {M}undprecht, {Z}aussinger, and
  {S}chiansky}]{Grimm-StreleKupkaLoew-Basellietal2015}
\bibinfo{author}{H.~{Grimm-Strele}}, \bibinfo{author}{F.~{K}upka},
  \bibinfo{author}{B.~{L\"ow-Baselli}}, \bibinfo{author}{E.~{M}undprecht},
  \bibinfo{author}{F.~{Z}aussinger}, \bibinfo{author}{P.~{S}chiansky},
\newblock \bibinfo{journal}{NewA} \bibinfo{volume}{34} (\bibinfo{year}{2015})
  \bibinfo{pages}{278 -- 293}.
\bibitem[{{R}ogers et~al.(1996){R}ogers, {S}wenson, and
  {I}glesias}]{RogersSwensonIglesias1996}
\bibinfo{author}{F.~J. {R}ogers}, \bibinfo{author}{F.~J. {S}wenson},
  \bibinfo{author}{C.~A. {I}glesias},
\newblock \bibinfo{journal}{ApJ} \bibinfo{volume}{456} (\bibinfo{year}{1996})
  \bibinfo{pages}{902}.
\bibitem[{{Kurucz}(1993{\natexlab{a}})}]{Kurucz1993CD13}
\bibinfo{author}{R.~{Kurucz}},
\newblock \bibinfo{journal}{Kurucz CD-ROM No.~13.~ Cambridge, Mass.:
  Smithsonian Astrophysical Observatory}  (\bibinfo{year}{1993}{\natexlab{a}}).
\bibitem[{{Kurucz}(1993{\natexlab{b}})}]{Kurucz1993CD2}
\bibinfo{author}{R.~{Kurucz}},
\newblock \bibinfo{journal}{Kurucz CD-ROM No.~2.~Cambridge, Mass.: Smithsonian
  Astrophysical Observatory}  (\bibinfo{year}{1993}{\natexlab{b}}).
\bibitem[{{I}glesias and {R}ogers(1996)}]{IglesiasRogers1996}
\bibinfo{author}{C.~A. {I}glesias}, \bibinfo{author}{F.~J. {R}ogers},
\newblock \bibinfo{journal}{ApJ} \bibinfo{volume}{464} (\bibinfo{year}{1996})
  \bibinfo{pages}{943}.
\bibitem[{{Grevesse} and {Noels}(1993)}]{GrevesseNoels1993}
\bibinfo{author}{N.~{Grevesse}}, \bibinfo{author}{A.~{Noels}},
\newblock in: \bibinfo{editor}{N.~{Prantzos}},
  \bibinfo{editor}{E.~{Vangioni-Flam}}, \bibinfo{editor}{M.~{Casse}} (Eds.),
  \bibinfo{booktitle}{Origin and Evolution of the Elements},
  \bibinfo{year}{1993}, pp. \bibinfo{pages}{15 -- 25}.
\bibitem[{{T}aylor et~al.(2007){T}aylor, {W}u, and
  {M}artin}]{TaylorWuMartin2007}
\bibinfo{author}{E.~M. {T}aylor}, \bibinfo{author}{M.~{W}u},
  \bibinfo{author}{M.~P. {M}artin},
\newblock \bibinfo{journal}{JCP} \bibinfo{volume}{223} (\bibinfo{year}{2007})
  \bibinfo{pages}{384 -- 397}.
\bibitem[{{S}tein and {N}ordlund(2000)}]{SteinNordlund2000}
\bibinfo{author}{R.~F. {S}tein}, \bibinfo{author}{A.~{N}ordlund},
\newblock \bibinfo{journal}{Solar Physics} \bibinfo{volume}{192}
  (\bibinfo{year}{2000}) \bibinfo{pages}{91 -- 108}.
\bibitem[{{K}upka(2009)}]{Kupka2009b}
\bibinfo{author}{F.~{K}upka}, \bibinfo{title}{{T}urbulent {C}onvection and
  {S}imulations in {A}strophysics}, \bibinfo{publisher}{Springer Lecture Notes
  in Physics 756}, \bibinfo{year}{2009}, pp. \bibinfo{pages}{49--105}.

\end{thebibliography}


\end{document}